\documentclass[journal]{IEEEtran}
\usepackage[utf8]{inputenc}

\title{Concatenated Forward Error Correction with \\ KP4 and Single Parity Check Codes}
\author{Diego Lentner, Emna Ben Yacoub, Stefano Calabr{\`o}, Georg B{\"o}cherer, Neboj\v{s}a Stojanovi{\'c}, Gerhard Kramer%
\thanks{D. Lentner, E. Ben Yacoub, and G. Kramer are with the Institute for Communications Engineering, School of Computation, Information and Technology, Technical University of Munich, 80333 Munich, Germany. E-mail: \{diego.lentner, emna.ben-yacoub, gerhard.kramer\}@tum.de.}%
\thanks{ S. Calabr{\`o}, G. B{\"o}cherer, and N. Stojanovi{\'c} are with the Huawei Munich Research Center, 80992 Munich, Germany. E-mail: \{stefano.calabro, georg.bocherer, nebojsa.stojanovic\}@huawei.com.}}

\usepackage{cite}
\usepackage{graphicx}
\usepackage{TUMcolors}
\usepackage[nolist]{acronym}
\usepackage{csquotes}
\usepackage{mathtools}

\usepackage{amsmath}
\usepackage{amssymb}
\usepackage{tikz}
\usepackage{pgfplots}
\usepackage[detect-all,binary-units]{siunitx}

\usepackage[hyphens]{url}
\usepackage{hyperref}
\usepackage[hyphenbreaks]{breakurl}
\usepackage{bm}

\allowdisplaybreaks

\usepackage[caption=false,font=footnotesize]{subfig}

\usepackage{amsthm}
\theoremstyle{definition}
\newtheorem{exmp}{Example}%

\begin{acronym}
    \acro{2D-SPC}{two-dimensional product code with SPC component codes}
    \acro{CDF}{cumulative distribution function}
	\acro{ASK}{amplitude shift keying}
	\acro{AWGN}{additive white Gaussian noise}
	\acro{BCH}{Bose–Chaudhuri–Hocquenghem}
	\acro{BDD}{bounded distance decoding}
	\acro{BER}{bit error rate}
	\acro{BiAWGN}{binary-input additive white Gaussian noise}
	\acro{BICM}{bit-interleaved coded modulation}
	\acro{BMD}{bit-metric decoding}
	\acro{BPSK}{binary phase shift keying}
	\acro{BSC}{binary symmetric channel}
	\acro{CatFEC}{concatenated FEC}
	\acro{DCI}{data center interconnect}
	\acro{DCN}{data center network}
	\acro{FEC}{forward error correction}
	\acro{FER}{frame error rate}
	\acro{HDD}{hard-decision decoding}
	\acro{IC}{integrated circuit}
	\acro{iid}{independent and identically distributed}
	\acro{ISI}{inter-symbol interference}
	\acro{LUT}{look-up table}
	\acro{LDPC}{low-density parity-check}
	\acro{LLR}{log-likelihood ratio}
	\acro{LMMSE}{linear minimum mean squared error}
	\acro{LRB}{least reliable bit}
	\acro{MAP}{maximum a posteriori}
	\acro{MI}{mutual information}
	\acro{ML}{maximum likelihood}
	\acro{MLC}{multilevel coding}
	\acro{MSD}{multi-stage decoding}
	\acro{NCG}{net coding gain}
	\acro{OH}{overhead}
	\acro{PAM}{pulse amplitude modulation}
	\acro{PMF}{probability mass function}
	\acro{QAM}{quadrature amplitude modulation}
	\acro{RS}{Reed-Solomon}
	\acro{SE}{spectral efficiency}
	\acro{SerDes}{serializer/deserializer transceiver}
	\acro{SC}{successive cancellation}
	\acro{SDD}{soft-decision decoding}
	\acro{SCL}{successive cancellation list}
	\acro{SMD}{symbol-metric decoding}
	\acro{SPC}{single parity check}
	\acro{SNR}{signal-to-noise ratio}
	\acro{QMP}{quaternary message passing}
\end{acronym}

\DeclareMathOperator{\sgn}{sgn}
\newcommand{\diff}{\ensuremath{\mathrm{d}}}
\providecommand{\abs}[1]{\ensuremath{\left\lvert#1\right\rvert}}

\renewcommand{\Pr}[1]{{\rm Pr}\ensuremath{\left[#1\right]}}

\newcommand{\kspc}{\ensuremath{k_\mathsf{SPC}}}
\newcommand{\nspc}{\ensuremath{n_\mathsf{SPC}}}

\newcommand{\kh}{\ensuremath{k_\mathsf{H}}}
\newcommand{\nh}{\ensuremath{n_\mathsf{H}}}

\newcommand{\krs}{\ensuremath{k_\mathsf{RS}}}
\newcommand{\nrs}{\ensuremath{n_\mathsf{RS}}}

\newcommand{\krsb}{\ensuremath{\tilde{k}_{\mathsf{RS}}}}
\newcommand{\nrsb}{\ensuremath{\tilde{n}_{\mathsf{RS}}}}

\newcommand{\EbNo}{\ensuremath{\mathsf{E_b}/\mathsf{N_0}}}
\newcommand{\SNR}{\ensuremath{\mathsf{SNR}}}
\newcommand{\BERin}{\ensuremath{\mathsf{BER}_{\mathsf{in}}}}

\newcommand{\Pf}{\ensuremath{\mathsf{P_f}}}
\newcommand{\Pb}{\ensuremath{\mathsf{P_b}}}
\newcommand{\pt}{\ensuremath{\tilde{p}}}
\newcommand{\Pfrs}{\ensuremath{\mathsf{P_{f,RS}}}}
\newcommand{\Psrs}{\ensuremath{\mathsf{P_{s,RS}}}}
\newcommand{\Pbrs}{\ensuremath{\mathsf{P_{b,RS}}}}
\newcommand{\Pfspc}{\ensuremath{\mathsf{P_{f,SPC}}}}
\newcommand{\Pbspc}{\ensuremath{\mathsf{P_{b,SPC}}}}

\newcommand{\pia}{\ensuremath{\mathsf{\pi}}}
\newcommand{\ipia}{\ensuremath{\mathsf{\pi^{-1}}}}
\newcommand{\npia}{\ensuremath{\abs{\pia}}}

\usetikzlibrary{calc}

\begin{document}

\maketitle

\begin{abstract}
Concatenated forward error correction is studied using an outer KP4 Reed-Solomon code with hard-decision decoding and inner single parity check (SPC) codes with Chase/Wagner soft-decision decoding. Analytical expressions are derived for the end-to-end frame and bit error rates for transmission over additive white Gaussian noise channels with binary phase-shift keying (BPSK) and quaternary amplitude shift keying (4-ASK), as well as with symbol interleavers and quantized channel outputs. The BPSK error rates are compared to those of two other inner codes: a two-dimensional product code with SPC component codes and an extended Hamming code. Simulation results for unit-memory inter-symbol interference channels and 4-ASK are also presented. The results show that the coding schemes achieve similar error rates, but SPC codes have the lowest complexity and permit flexible rate adaptation.
\end{abstract}

\section{Introduction}
\IEEEPARstart{A}{pplications} such as \acp{DCN} and \acp{DCI} over short-reach fiber optic links have high throughput and strict latency constraints. Next-generation Ethernet standards should provide data rates of $\SI{800}{Gb/s}$ to $\SI{1.6}{Tb/s}$ \cite{dambrosia2021project}, \cite{800GbE} while guaranteeing \acp{BER} below $10^{-13}$ and latencies below $\SI{100}{ns}$ \cite{he2021fec}, \cite{dambrosia2022objectives}. \Ac{FEC} is essential to meet these requirements, and current systems rely on the KP4 code \cite{ieee802.3}, which is a $(544,514)$ \ac{RS} code over $\mathbb{F}_{2^{10}}$ with the rate $R_{\mathsf{KP4}}\approx 0.945$ and \ac{OH}
$\approx 5.84\%$. 

The upcoming 800GbE standard specifies a $\SI{800}{Gb/s}$ data stream supported by four $\SI{200}{Gb/s}$ optical signals which can, e.g., be transmitted over four parallel single-mode (PSM) fibers or four wavelengths on the same fiber in coarse division wavelength multiplexing.
The transition from $\SI{100}{Gb/s}$ to $\SI{200}{Gb/s}$ per optical signal motivates improving the KP4 code by using \ac{CatFEC}~\cite{forney65thesis}, which has been used for medium-range links~\cite{400g-zr} where the latency constraints are less strict.
Our work is also motivated by hardware solutions where signal processing is done on pluggable or co-packaged optical modules at both ends of the optical link and by placing a low-complexity inner \ac{FEC} decoder on these modules; see~\cite{dambrosia2021project}, \cite{he2021fec}. In this setup, the inner decoder can access quantized channel measurements that may be used, e.g., for equalization and \ac{SDD} of the inner code. The outer KP4 decoder is usually implemented on the Ethernet switch chip and accepts hard decisions as input.

We study \ac{CatFEC} with an outer KP4 code and inner \ac{SPC} codes. \ac{SPC} codes have a simple \ac{SDD} algorithm \cite{silverman1954}, and their rate can be adapted with small granularity by varying the code length, permitting flexible rate adaptation for channels of varying qualities. In the magnetic recording literature, the paper \cite{chaichanavong2007on} studied \ac{CatFEC} with an outer \ac{RS} code and inner \ac{SPC} codes with \ac{SDD}, and the authors provided semi-analytical methods to estimate the end-to-end error probabilities. We extend their work by deriving analytical expressions for the end-to-end \acp{FER} and \acp{BER} over \ac{AWGN} channels. This permits performance assessment without using numerical simulations. We compare the performance to solutions with symbol interleavers and where the inner code is either a \ac{2D-SPC} or an extended Hamming code \cite{nedelcu2022concatenated}.

This paper is organized as follows. Section~\ref{ch:CFEC} introduces the proposed \ac{CatFEC} scheme. Section~\ref{ch:analysis} derives analytical expressions for the end-to-end \acp{FER} and \acp{BER} with \ac{BPSK}, both without and with symbol interleavers. Sections~\ref{sec:analysis:coarslyquantized} and~\ref{sec:analysis:higherorder} extend the analysis to coarsely quantized inputs and to quaternary amplitude shift keying (4-ASK). Section~\ref{ch:comparison} compares the proposed \ac{CatFEC} scheme to two alternative methods using a \ac{2D-SPC} inner code and an extended Hamming inner code. Section~\ref{sec:ISI_channel} provides simulation results for 4-ASK and a unit-memory \ac{ISI} channel. Section~\ref{ch:conclusion} concludes the paper.
        
\section{Concatenated Scheme}
\label{ch:CFEC}

\begin{figure}
    \centering
    \scalebox{0.7}{\includegraphics{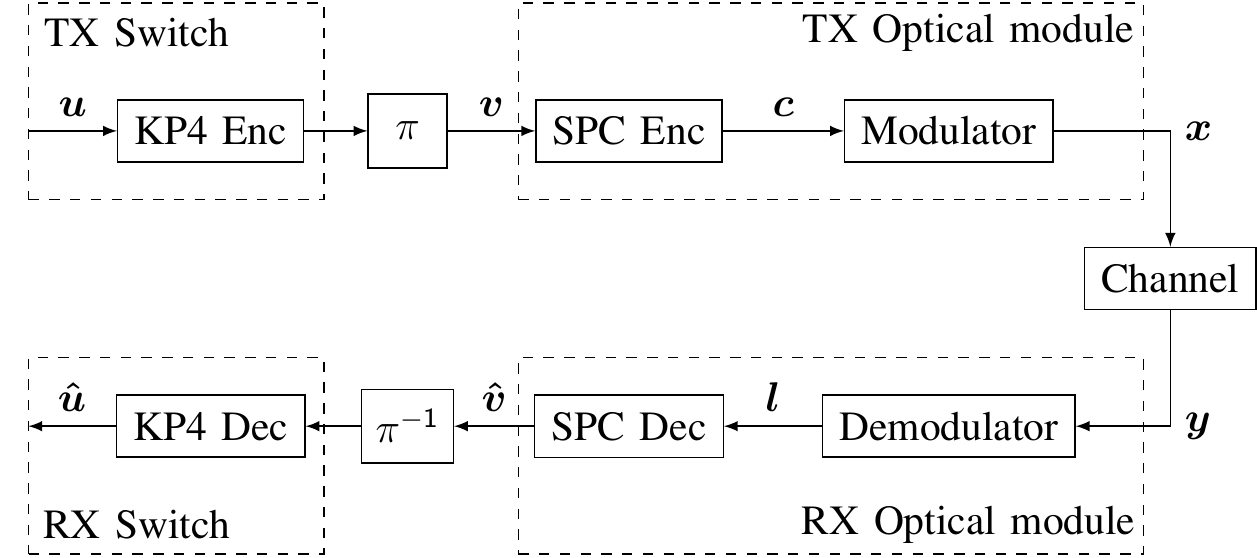}}
    \caption{\ac{CatFEC} with an outer KP4 code and an inner SPC code with \ac{SDD}. }
    \label{fig:concat_diag}
\end{figure}

The transceiver chain is depicted in Fig.~\ref{fig:concat_diag}. The switch chip has an outer KP4 code that is widely deployed in current systems, and we augment the optical modules with an inner \ac{SPC} code layer. We place a \ac{RS} code symbol interleaver $\pia$ of length $\npia$ between the outer and inner encoders to counteract error bursts of the inner decoder. Note that $\pia$ can be implemented on the switch \ac{IC}, or on the optical module, or partially on both.  In general, interleaving over multiple outer codewords increases decoding latency as the receiver must collect more code symbols before outer decoding. However, if \ac{FEC} is implemented jointly over $T$ optical signals, interleaving over $T$ outer codewords does not increase the system latency compared to implementing separate \ac{FEC} per optical signal without interleaving. 

We thus analyze two cases:
\begin{enumerate}
    \item No interleaver to model separate \ac{FEC} per optical signal. In this case, $\pia$ is the identity mapping.
    \item A uniform block-to-block symbol interleaver\footnote{``Uniform interleaver'' refers to an interleaver drawn with equal probability from the set of all permutations.} $\pia$ to model joint \ac{FEC} across multiple optical signals.
\end{enumerate}
We study uniform block-to-block interleavers for two reasons: to obtain analytical expressions for the error rates and because the analysis guarantees the existence of an interleaver that can achieve these error rates. Convolutional interleavers \cite{ramsey1970realization}, \cite{forney1971burst} might exhibit better latency vs.\ error-rate tradeoffs than block-to-block interleavers. 
Also, when the channel has memory, the system performance can be improved by inserting a second interleaver-deinterleaver pair between the modulator/demodulator and the channel.

Let $\boldsymbol{u}\in\mathbb{F}_2^{\krsb}$, $\krsb=5140$, be the binary input to the outer KP4 \ac{RS} encoder at the transmitter. After interleaving, the \ac{RS} codeword's $\nrsb=5440$ bits are grouped into blocks $\boldsymbol{v}\in\mathbb{F}_2^{\kspc}$ with $\kspc$ bits that are fed to the \ac{SPC} encoder. The \ac{SPC} encoder appends a parity bit, yielding the \ac{SPC} codeword $\boldsymbol{c}\in\mathbb{F}_2^{\nspc}$ where $\nspc=\kspc+1$. Finally, suppose the signal constellation $\mathcal{X}$ has cardinality $|\mathcal{X}|=2^K$ where $K$ is a positive integer. The code bits are grouped into blocks of $K$ bits and the modulator maps these blocks to symbols $x\in\mathcal{X}$ using the inverse of the labeling function $\chi:\mathcal{X}\to\mathbb{F}_2^K$. The $k$-th bit of $\chi(x)$ is written as $\chi_k(x)$, $k=1,\dots,K$.

At the receiver, the demodulator observes the channel output vector $\boldsymbol{y}$ with $\lceil\frac{1}{K}\cdot\nrsb\cdot\frac{\nspc}{\kspc}\rceil$ real values and converts these to the \ac{LLR} vector $\boldsymbol{l}$ with $\lceil\nrsb\cdot\frac{\nspc}{\kspc}\rceil$ real values.
Each \ac{SPC} codeword is then decoded individually using an \ac{SDD} algorithm, and multiple \ac{SPC} decoders can run in parallel to lower the decoding latency. The receiver optical module passes the hard decisions $\boldsymbol{\hat{v}}\in\mathbb{F}_2^{\kspc}$ of the systematic bits to the receiver switch module. Finally, the switch module applies deinterleaving and KP4 \ac{RS} decoding to output the final estimate $\hat{\boldsymbol{u}}\in\mathbb{F}_2^{\krsb}$.

A low-complexity \ac{SDD} algorithm for \ac{SPC} codes is Wagner decoding \cite{silverman1954}. This algorithm performs bitwise \ac{HDD} with $\tilde{c}_i = \chi(\sgn({l_i}))$,
$i=1,\ldots,\nspc$, and checks if the parity check constraint is satisfied. If $\boldsymbol{\tilde{c}}$ is a valid \ac{SPC} codeword, then the decoder outputs the hard decisions of the systematic bits $\boldsymbol{\hat{v}}=\begin{bmatrix} \tilde{c}_1 & \ldots & \tilde{c}_{\kspc} \end{bmatrix}$. Otherwise, it finds the \ac{LRB} position\footnote{There may be multiple positions $i'$ of smallest reliability when the \acp{LLR} are quantized. The decoder may choose any of the \ac{LRB} positions in this case.} $i' = \arg\min_i \abs{l_i}$, flips the corresponding bit
\begin{align}
    \boldsymbol{\hat{c}} = \begin{bmatrix} 
    \tilde{c}_1 & \ldots & \tilde{c}_{i'}\oplus 1 & \ldots & \tilde{c}_{\nspc}  \end{bmatrix}
\end{align}
and outputs the systematic part $\boldsymbol{\hat{v}} = \begin{bmatrix} \hat{c}_1 & \ldots & \hat{c}_{\kspc} \end{bmatrix}$.
A Wagner decoder is a maximum-likelihood decoder for \ac{SPC} codes over \ac{AWGN} channels \cite{wolf1978} and an efficient implementation of Chase decoding \cite{chase1972}. Wagner decoding uses \ac{SDD} only to identify the \ac{LRB} but does not perform further computations, so the proposed \ac{CatFEC} scheme does not require a full-precision \ac{LLR} computation at the demodulator in general.

The end-to-end transmission rate of the \ac{CatFEC} scheme is
\begin{align}
    R = R_{\mathsf{KP4}} \cdot \frac{\nspc-1}{\nspc} \cdot \log_2 \abs{\mathcal{X}}
\end{align}
which can be adapted by varying $\nspc$. We do not require that $\nspc-1$ is an integer multiple of the outer RS field size, nor must $\nspc-1$ divide the outer code length. 

\begin{figure}
    \centering
    \includegraphics{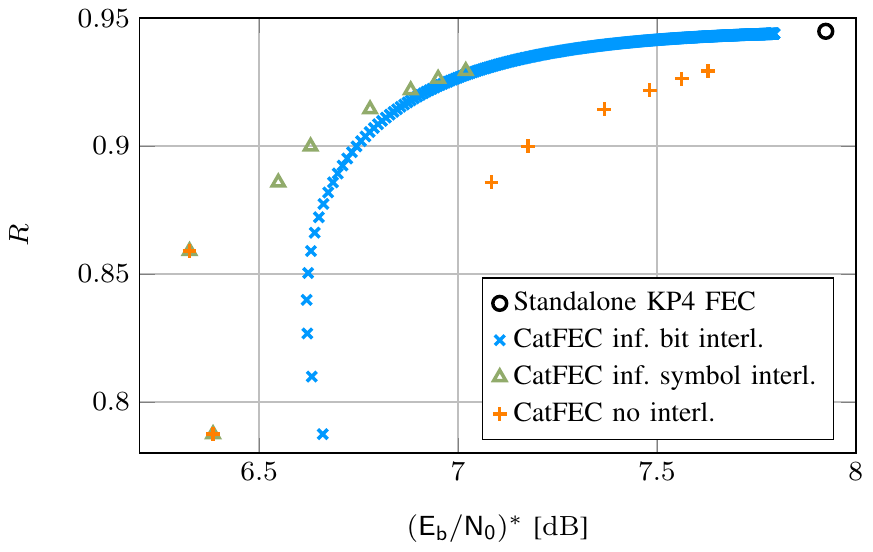}
    \caption{\ac{CatFEC} rate vs. $\EbNo$ threshold for \ac{BPSK} transmission over the \ac{AWGN} channel and an end-to-end BER of $10^{-13}$.}
    \label{fig:rates_vs_ebno}
\end{figure}

Fig.~\ref{fig:rates_vs_ebno} illustrates the fine granularity of rates permitted by SPC codes of different lengths. The plot is for transmission with \ac{BPSK} over an \ac{AWGN} channel. Let $\SNR^*$ be the \ac{SNR} threshold for which the \ac{CatFEC} schemes achieve the target end-to-end \ac{BER} of $10^{-13}$, and define $(\EbNo)^* = \SNR^*/(2 R)$. The KP4 threshold is indicated by a black circle. The blue markers show the rates for $\nspc=6,7,8,\ldots,1000$ for which the \ac{BER} after \ac{SPC} decoding meets the KP4 threshold of $3.1\times10^{-4}$. This \ac{BER} corresponds to the target \ac{BER} of $10^{-13}$ when the bit errors after \ac{SPC} decoding are independent, but this generally requires a bit interleaver of infinite length.
To compare, the orange crosses in Fig.~\ref{fig:rates_vs_ebno} show the rates without interleaving while the green triangles show the rates with an infinitely long \ac{RS} symbol interleaver, in both cases for $\nspc=6,11,16,21,31,\ldots,61$. Observe that a symbol interleaver performs slightly better than the bit interleaver. Also, for $\nspc=6,11$, the \ac{RS} symbol errors after \ac{SPC} decoding are independent, and no interleaving performs as well as symbol interleaving. The highest coding gain is achieved for $\nspc=11$, i.e., when one \ac{SPC} codeword covers exactly one \ac{RS} symbol.
The following section gives closed-form expressions for these error rates and the error rates with symbol interleaving.

\section{Error Analysis for BPSK}
\label{ch:analysis}

We use uppercase letters (such as $Y$) to denote random variables and lowercase letters (such as $y$) to denote realizations. Consider the \ac{AWGN} channel with output
\begin{align}
    Y_j = X_j + N_j
\end{align}
at time $j$, where $X_j\in\mathcal{X}$ and $N_j\sim\mathcal{N}(0,\sigma^2)$. 
This section studies \ac{BPSK} with $\mathcal{X}=\{+1,-1\}$ and the labeling function $\chi(+1)=0$ and $\chi({-1})=1$.
With slight abuse of terminology, we sometimes refer to the code bits as channel inputs. By symmetry, we may assume that $\boldsymbol{u}$ is the all-zeros string, so the marginal probability density function of the channel output is
\begin{align}
    p_Y(y)= \frac{1}{\sqrt{2 \pi \sigma^2}} e^{-\frac{(y-1)^2}{2 \sigma^2}} \,.
\end{align}
The channel \ac{LLR} $L=2y/\sigma^2$ is Gaussian distributed with mean $\mu_\text{ch}= 2/\sigma^2$ and variance $\sigma^2_\text{ch}=4/\sigma^2$.

The end-to-end \ac{FER} is $\Pf = \Pr{\hat{\boldsymbol{U}}\neq\boldsymbol{u}}$
and the average end-to-end \ac{BER} is
\begin{align}
    \Pb = \frac{1}{\krsb} \sum_{i=1}^{\krsb} \Pr{\hat{U}_i\neq u_i}.
\end{align}
Define $\SNR=1/\sigma^2$ where $\sigma^2$ is the noise variance per real dimension. We plot error rates against $\EbNo = \frac{\SNR}{2 R}$ to measure coding gains. Sometimes we plot error rates against the (uncoded) input $\BERin = Q(\sqrt{\SNR})$, where $Q(x)=\frac{1}{\sqrt{2 \pi}} \int_x^{\infty} e^{-\frac{a^{2}}{2}} \diff a$.

\subsection{Standalone RS Codes}
\label{sec:analysis:RS_standalone}
Consider a (perhaps shortened) \ac{RS} code of length $\nrs$ and dimension $\krs$ defined over $\mathbb{F}_{2^m}$. We assume \ac{BDD} so the decoder corrects any error pattern with $t=\lfloor\frac{\nrs-\krs}{2}\rfloor$ or fewer errors. 
For the KP4 code, we have $\nrs=544$, $\krs=514$, $m=10$, and $t=15$.
Let $p$ denote the \ac{BER} before \ac{RS} decoding. If the transmitted bits are independent and the channel is memoryless, we can express the uncoded RS symbol error probability at the channel output as $\pt = 1-(1-p)^m$. 
The output \ac{FER} and symbol error probability are
\begin{align}
    \Pfrs &= \sum_{i=t+1}^{\nrs} \binom{\nrs}{i} \pt^i (1-\pt)^{\nrs-i}
    \label{eq:Pfrs} \\
    \Psrs &= \frac{1}{\nrs} \sum_{i=t+1}^{\nrs} i \binom{\nrs}{i} \pt^i (1-\pt)^{\nrs-i}
    \label{eq:Psrs}
\end{align}
assuming that the decoder never miscorrects and outputs the input sequence in case of failure.

The average \ac{BER} is
\begin{align}
    \Pbrs &= \frac{p\cdot\Psrs}{\pt} \label{eq:Pbrs}
\end{align}
which can be well-approximated as $\Pbrs\approx\Psrs/m$ for small values of $p$ if the decoder input bits are independent.

\subsection{Standalone SPC Codes with Wagner Decoding}
\label{sec:analysis:SPC_standalone}
Let $\mathcal{E}$ be the event that Wagner decoding of an \ac{SPC} code fails. For $\ell=1,\ldots,\nspc$, let $\mathcal{A}_\ell$ be the event that $\ell$ of the $\nspc$ hard decisions at the channel output are received in error. Moreover, let $\mathcal{C}_1$ be the event that there is one erroneous bit, the \ac{LRB}, i.e., the decoder can successfully correct the only error. We have $\mathcal{E} = \lbrace\bigcup_{\ell=1}^{\nspc} \mathcal{A}_\ell\rbrace\setminus\mathcal{C}_1$ and the \ac{FER} of the \ac{SPC} code is
\begin{align}
    \Pfspc &=  \left( \sum_{\ell=1}^{\nspc} \Pr{\mathcal{A}_\ell} \right) - \Pr{\mathcal{C}_1}
    \label{eq:Pfspc}
\end{align} 
where
\begin{align}
    \Pr{\mathcal{A}_\ell} &= \binom{\nspc}{\ell}  p^\ell (1-p)^{\nspc-\ell} \label{eq:Pr_Al}
\end{align}
with $p=\BERin=Q(1/\sigma)$. To compute $\Pr{\mathcal{C}_1}$, by symmetry we may assume that $c_1$ is the \ac{LRB} and write
\begin{align}
    \Pr{\mathcal{C}_1}
    & = \nspc \, \Pr{ \{Y_1 < 0\} \cap \left\{\bigcap_{j=2}^{\nspc} \left\{Y_j \ge |Y_1|\right\} \right\} } \nonumber \\
    &= \nspc \int\limits_{-\infty}^{0} p_Y(y) \; Q\left( \frac{-y-1}{\sigma}\right)^{\nspc-1} \diff y  
\end{align}
which can be easily evaluated by numerical integration.

If one is interested only in the \ac{FER}, the computation of $\Pfspc$ can be simplified as follows. Let $\mathcal{A}_0$ be the event that all input bits are received correctly so that $\Pr{\mathcal{A}_0} = (1-p)^{\nspc}$ and $\Pr{\bigcup_{\ell=0}^{\nspc} \mathcal{A}_\ell}=1$. It follows that
$\sum_{\ell=1}^{\nspc} \Pr{\mathcal{A}_\ell} = 1-(1-p)^{\nspc}$. Inserting this result into \eqref{eq:Pfspc} gives
\begin{align}
    \Pfspc &= 1-(1-p)^{\nspc} - \Pr{\mathcal{C}_1}
    \label{eq:Pfspc_simplified}
\end{align} 
without evaluating \eqref{eq:Pr_Al} for each  $\ell=1,\ldots,\nspc$. If one is also interested in the \ac{BER}, however, $\Pr{\mathcal{A}_\ell}$ has to be computed anyways, as we shall see next.

To compute the \ac{BER} $\Pbspc$, we must determine how Wagner decoding affects the number of erroneous input bits. For \ac{BPSK}, all code bits have the same reliability after Wagner decoding, i.e., the \ac{BER} is independent of the bit position. This lets us avoid making case distinctions with respect to the parity bit that is later discarded.

Suppose the input to the decoder has $\ell$ bit errors. If $\ell$ is even, then after decoding the codeword will still have $\ell$ bit errors; the probability of this event is $\Pr{\mathcal{A}_\ell}$. If $\ell$ is odd, we distinguish two cases: the correction event $\mathcal{C}_{\ell}$ that the \ac{LRB} is among the erroneous bits and the miscorrection event $\mathcal{M}_{\ell}$ that the \ac{LRB} is not among the erroneous bits. The average \ac{BER} after Wagner decoding measured with respect to the code bits or the information bits is then
\begin{align}
    \Pbspc &= \sum_{\substack{\ell=2\\ \ell \text{ even}}}^{\nspc} \frac{\ell}{\nspc} \cdot \Pr{\mathcal{A}_\ell}
    + \sum_{\substack{\ell=3\\ \ell \text{ odd}}}^{\nspc} \frac{\ell-1}{\nspc} \cdot \Pr{\mathcal{C}_\ell} \nonumber\\
    &\quad+ \sum_{\substack{\ell=1\\ \ell \text{ odd}}}^{\nspc-1} \frac{\ell+1}{\nspc} \cdot \Pr{\mathcal{M}_\ell}
    \label{eq:Pbspc}
\end{align}
where for $\ell=1,\ldots,\nspc$ we have
\begin{align}
    \Pr{\mathcal{C}_\ell} &= \binom{\nspc}{\ell} \cdot \ell \cdot \int\limits_{-\infty}^{0} \phi_\ell(y) \,\diff y  \\
    \Pr{\mathcal{M}_\ell} &= \binom{\nspc}{\ell} \cdot (\nspc-\ell) \cdot \int\limits_{0}^{\infty} \psi_\ell(y) \,\diff y 
\end{align}
with the respective
\begin{align}
    \phi_\ell(y) &= p_Y(y) \; Q\left( \tfrac{-y+1}{\sigma}\right)^{\ell-1} Q\left( \tfrac{-y-1}{\sigma}\right)^{\nspc-\ell} \\
    \psi_\ell(y) &= p_Y(y) \; Q\left( \tfrac{y+1}{\sigma}\right)^{\ell} Q\left( \tfrac{y-1}{\sigma}\right)^{\nspc-\ell-1}.
\end{align}
Note that Wagner decoding outputs an even number of errors, as reflected in \eqref{eq:Pbspc}. Using $\Pr{\mathcal{A}_\ell}=\Pr{\mathcal{C}_\ell}+\Pr{\mathcal{M}_\ell}$, we can rewrite \eqref{eq:Pbspc} as
\begin{align}
    \Pbspc &= \frac{1}{\nspc} \left(2\cdot \Pr{\mathcal{M}_1} + \sum_{\ell=2}^{\nspc} \ell\cdot\Pr{\mathcal{A}_\ell} \right) \nonumber \\
    & \quad + \sum_{\substack{\ell=3\\ \ell \text{ odd}}}^{\nspc} \frac{1}{\nspc} \left( \Pr{\mathcal{M}_\ell}  - \Pr{\mathcal{C}_\ell} \right) .
    \label{eq:Pbspc1}
\end{align}
We use the exact expressions \eqref{eq:Pbspc} or \eqref{eq:Pbspc1} for our results. As a simplification, simulations show that a good approximation and upper bound for $\BERin< 10^{-2}$  is
\begin{align}
    \Pbspc &\approx \frac{1}{\nspc} \left(2\cdot \Pr{\mathcal{M}_1 } + \sum_{\ell=2}^{\nspc} \ell\cdot\Pr{\mathcal{A}_\ell} \right).
    \label{eq:Pbspc_approx}
\end{align}

\subsection{Refined Analysis}
\label{sec:analysis:spc_refined}
We are interested in the probability that all bit errors are in one part of a codeword. Let $\ell$ again denote the number of erroneous code bits at the input of the \ac{SPC} decoder. By symmetry, we may consider the event  $\mathcal{\widehat{A}}_\ell^{\kappa+1}$ that the first $\nspc-\kappa-1$ code bits are error-free \emph{before} decoding and that all $\ell$ input bit errors are in the last $\kappa+1$ code bits that include the parity bit. We compute
\begin{align}
    &\Pr{\mathcal{\widehat{A}}_\ell^{\kappa+1}} = \binom{\kappa+1}{\ell}  p^\ell (1-p)^{\nspc-\ell}. \label{eq:Pr_Al_part}
\end{align}
Similarly, let $\mathcal{\widehat{E}}_\ell^{\kappa+1}$ be the event that the first $\nspc-\kappa-1$ code bits are error-free \emph{after} decoding, i.e., all bit errors after decoding are in the last $\kappa+1$ bits that include the parity bit that is discarded. For even $\ell$, we have $\mathcal{\widehat{E}}_\ell^{\kappa+1} = \mathcal{\widehat{A}}_\ell^{\kappa+1}$. For odd $\ell$, we distinguish the following disjoint events:
\begin{itemize}
\item the correction event
\begin{align*}
    \mathcal{\widehat{C}}_\ell^{\kappa+1} = \mathcal{\widehat{A}}_\ell^{\kappa+1} \cap \{\text{\ac{LRB} is erroneous}\};
\end{align*}
\item the miscorrection event \begin{align*}
    \mathcal{\widehat{M}}_\ell^{\kappa+1} & = \mathcal{\widehat{A}}_\ell^{\kappa+1} \cap \{\text{\ac{LRB} is not erroneous}\} \\ & \quad \cap \{\text{\ac{LRB} is in the last $\kappa+1$ code bits}\};
\end{align*}
\item the cross-miscorrection event \begin{align*}
    \mathcal{\widehat{H}}_\ell^{\kappa+1} & = \mathcal{\widehat{A}}_\ell^{\kappa+1} \cap \{\text{\ac{LRB} is not erroneous}\} \\ & \quad \cap \{\text{\ac{LRB} is not in the last $\kappa+1$ code bits}\};
\end{align*}
\item the cross-correction event $\mathcal{\widehat{K}}_\ell^{\kappa+1}$ that only $\ell-1$ errors are in the last $\kappa+1$ code bits before decoding, and the erroneous \ac{LRB} that is corrected was in the first $\nspc-\kappa-1$ code bits. 
\end{itemize}
The respective probabilities of these events are
\begin{align}
    &\Pr{\mathcal{\widehat{C}}_\ell^{\kappa+1}} = \binom{\kappa+1}{\ell} \cdot \ell \cdot \int\limits_{-\infty}^{0} \phi_\ell(y) \,\diff y \label{eq:Pr_Cl_part}\\
    &\Pr{\mathcal{\widehat{M}}_\ell^{\kappa+1}} = \binom{\kappa+1}{\ell} \cdot (\kappa+1-\ell) \cdot \int\limits_{0}^{\infty} \psi_\ell(y) \,\diff y \label{eq:Pr_Ml_part}\\
    &\Pr{\mathcal{\widehat{H}}_\ell^{\kappa+1}} = \binom{\kappa+1}{\ell} \cdot (\nspc-\kappa-1) \cdot \int\limits_{0}^{\infty} \psi_\ell(y) \,\diff y \label{eq:Pr_Hl_part}\\
    &\Pr{\mathcal{\widehat{K}}_\ell^{\kappa+1}} = \binom{\kappa+1}{\ell-1} \cdot (\nspc-\kappa-1) \cdot \int\limits_{-\infty}^{0} \phi_\ell(y) \,\diff y . \label{eq:Pr_Kl_part}
\end{align}
We have $\mathcal{\widehat{A}}_\ell^{\kappa+1} = \mathcal{\widehat{C}}_\ell^{\kappa+1} \cup \mathcal{\widehat{M}}_\ell^{\kappa+1} \cup \mathcal{\widehat{H}}_\ell^{\kappa+1}$ and $\mathcal{\widehat{H}}_\ell^{\kappa+1} \cap \mathcal{\widehat{E}}_\ell^{\kappa+1}= \emptyset$.
We also have $\mathcal{\widehat{K}}_\ell^{\kappa+1} \cap \mathcal{\widehat{A}}_\ell^{\kappa+1} = \emptyset$ but $\mathcal{\widehat{K}}_\ell^{\kappa+1} \subseteq \mathcal{\widehat{E}}_\ell^{\kappa+1}$. We thus find for odd $\ell$ that
\begin{align*}
    \mathcal{\widehat{E}}_\ell^{\kappa+1} &= \mathcal{\widehat{K}}_\ell^{\kappa+1} \cup \lbrace\mathcal{\widehat{A}}_\ell^{\kappa+1} \setminus \mathcal{\widehat{H}}_\ell^{\kappa+1}\rbrace \\
    &= \mathcal{\widehat{K}}_\ell^{\kappa+1} \cup \mathcal{\widehat{C}}_\ell^{\kappa+1} \cup \mathcal{\widehat{M}}_\ell^{\kappa+1}.
\end{align*}
Note that also $\lbrace\mathcal{\widehat{K}}_\ell^{\kappa+1} \cup \mathcal{\widehat{A}}_\ell^{\kappa+1}\rbrace \subsetneq \mathcal{A}_\ell$ since the cross-miscorrections $\mathcal{\widehat{H}}_\ell^{\kappa+1}$ are not the only events that we exclude in our constrained error rate computation. For instance, we exclude all events where two or more of the $\ell$ input errors are in the first $\nspc-\kappa-1$ bits before decoding.

The constrained \ac{FER} with all bit errors in the last $\kappa+1$ code bits is
\begin{align}
    \mathsf{\hat{P}_{f,SPC}}(\kappa) &=  \sum_{\substack{\ell=2\\ \ell \text{ even}}}^{\kappa+1} \Pr{\mathcal{\widehat{A}}_\ell^{\kappa+1}}
    + \sum_{\substack{\ell=3\\ \ell \text{ odd}}}^{\kappa+1} \Pr{\mathcal{\widehat{C}}_\ell^{\kappa+1}} \nonumber\\
    &\quad+ \sum_{\substack{\ell=3\\ \ell \text{ odd}}}^{\kappa+2} \Pr{\mathcal{\widehat{K}}_\ell^{\kappa+1}}
    + \sum_{\substack{\ell=1\\ \ell \text{ odd}}}^{\kappa} \Pr{\mathcal{\widehat{M}}_\ell^{\kappa+1}} .\label{eq:Pfspc_part} 
\end{align} 
Unlike in \eqref{eq:Pfspc}, the constrained \ac{BER} of the systematic bits differs from the constrained \ac{BER} over the entire codeword, as the parity bit that is discarded may be erroneous. We therefore compute the constrained systematic \ac{BER} $\mathsf{\hat{P}_{b,SPC}}(\kappa)$ as
\begin{align}
    &\frac{\frac{\kappa}{\kappa+1}}{\kspc} \left( \sum_{\substack{\ell=2\\ \ell \text{ even}}}^{\kappa+1} \ell\cdot\Pr{\mathcal{\widehat{A}}_\ell^{\kappa+1}}
    + \sum_{\substack{\ell=3\\ \ell \text{ odd}}}^{\kappa+1} (\ell-1)\cdot \Pr{\mathcal{\widehat{C}}_\ell^{\kappa+1}} \right. \nonumber\\
    &\left. \quad+ \sum_{\substack{\ell=3\\ \ell \text{ odd}}}^{\kappa+2} (\ell-1)\cdot \Pr{\mathcal{\widehat{K}}_\ell^{\kappa+1}}
    + \sum_{\substack{\ell=1\\ \ell \text{ odd}}}^{\kappa} (\ell+1)\cdot \Pr{\mathcal{\widehat{M}}_\ell^{\kappa+1}} \right)
    \label{eq:Pbspc_part}
\end{align}
where the factor $\frac{\kappa}{\kappa+1}$ accounts for the fraction of bit errors in the systematic bits only. 
Note that $\mathsf{\hat{P}_{b,SPC}}(\kspc) = \Pfspc$, i.e., \eqref{eq:Pbspc_part} reduces to \eqref{eq:Pbspc} if we allow all code bits to be erroneous since $\mathcal{\widehat{K}}_\ell^{\kappa+1}=\emptyset$ and $\mathcal{\widehat{H}}_\ell^{\kappa+1}=\emptyset$ for all $\ell$.

\subsection{Error Rates Without Interleaver}
\label{sec:analysis:trivialinterleaver}
Suppose there is no interleaver between the outer and inner codes and let $\mathrm{LCM}(x,y)$ be the least common multiple of $x$ and $y$. The inner \ac{SPC} decoder causes burst errors across $\tau = \mathrm{LCM}(\kspc,m)/m$ 
outer \ac{RS} code symbols. If $\tau$ divides $\nrs$, then each RS codeword can be partitioned into $n_\tau = \frac{\nrs}{\tau}$ blocks of $\tau$ RS symbols that we study separately.
For $i=0,\ldots,\tau$, let $P_i$ be the probability that exactly $i$ RS symbols of the $\tau$-tuple are in error before outer decoding, and let $z_i$ be the total number of such $\tau$-tuples with $i$ erroneous symbols within one RS codeword. We have $\sum_{i=0}^{\tau} P_i = 1$, $\sum_{i=0}^{\tau} z_i= n_\tau$, and the end-to-end \ac{FER} is
\begin{align}
    \Pf = \sum\limits_{\substack{z_0,\ldots,z_{\tau}\,\geq 0 \\ \sum_{i=0}^{\tau} z_i \,=\, n_\tau \\  \sum_{i=1}^{\tau} i\cdot z_i \,\geq\, t+1}} \binom{n_\tau}{z_0,\,\ldots,\,z_{\tau}} \prod_{i=0}^{\tau} P_i^{z_i}
    \label{eq:Pf_general}
\end{align}
where $\binom{n}{k_0,\,\ldots,\,k_r} = \frac{n!}{k_0!\cdots k_r!}$ is the multinomial coefficient.
Furthermore, let $P_{b,i}$ be the \ac{BER} due to $\tau$-tuples with $i$ symbol errors before outer decoding such that $\sum_{i=1}^\tau P_{b,i}=\Pbspc$.  
The \ac{CatFEC} \ac{BER} is then
\begin{align}
    \Pb = \sum\limits_{\substack{z_0,\ldots,z_{\tau}\,\geq 0 \\ \sum_{i=0}^{\tau} z_i \,=\, n_\tau \\  \sum_{i=1}^{\tau} i\cdot z_i \,\geq\, t+1}}
    \left(\frac{\sum\limits_{i=1}^{\tau} z_i \frac{P_{b,i}}{P_i} }{n_\tau}\right)
    \binom{n_\tau}{z_0,\,\ldots,\,z_{\tau}} \prod_{i=0}^{\tau} P_i^{z_i} \,.
    \label{eq:Pb_general}
\end{align}
For $\tau=1$, \eqref{eq:Pf_general} and \eqref{eq:Pb_general} are equivalent to the basic \ac{RS} error probability expressions \eqref{eq:Pfrs}--\eqref{eq:Pbrs} with $\pt = P_1$ and $p=P_{b,1}$.

\begin{exmp}[KP4 \& inner $(11,10)$ SPC codes] \label{exmp:n11}
We have $\kspc = m$, i.e., each \ac{SPC} covers exactly one outer \ac{RS} symbol, which yields the case above where $\tau = 1$ and $n_\tau = \nrs = 544$. The \ac{FER} and \ac{BER} of the \ac{CatFEC} scheme are obtained by inserting $P_1= \Pfspc$ and $P_{b,1} = \Pbspc$ into \eqref{eq:Pf_general} and \eqref{eq:Pb_general}.
\end{exmp}

\begin{exmp}[KP4 \& inner $(6,5)$ SPC codes] \label{exmp:n6}
We again have $\tau = 1$ and $n_\tau = \nrs = 544$ since each outer RS symbol is protected by exactly two \ac{SPC} codes. This time, however, $P_1 = 1-(1-\Pfspc)^2$ and $P_{b,1} = \Pbspc$.
\end{exmp}

\begin{exmp}[KP4 \& inner $(21,20)$ SPC codes] \label{exmp:n21}
We have $\tau=2$ and $n_\tau = \nrs/2 = 272$, i.e., two \ac{RS} symbols are coupled by one inner \ac{SPC} code.
The probability that the first \ac{RS} symbol is error-free and the second symbol is erroneous is $\mathsf{\hat{P}_{f,SPC}}(10)$, and vice versa. We thus have $P_1 = 2\cdot\mathsf{\hat{P}_{f,SPC}}(10)$ and $P_2=\Pfspc-P_1$. Likewise, we have $P_{b,1} = 2\cdot\mathsf{\hat{P}_{b,SPC}}(10)$ and $P_{b,2}=\Pbspc-P_{b,1}$.
\end{exmp}

\begin{exmp}[KP4 \& inner $(\tau m+1,\tau m)$ SPC codes] \label{exmp:matched}
In this case, one \ac{SPC} codeword couples $\tau$ \ac{RS} symbols. Let $P_i'$, $i=1,\ldots,\tau$, be the probability that all the first $i$ \ac{RS} symbols of one $\tau$-tuple are erroneous and the remaining $\tau-i$ symbols are error-free. We have the recursion $P_i'= \mathsf{\hat{P}_{f,SPC}}(im) -\sum_{j=1}^{i-1}\binom{i}{j} P_j'$ and compute $P_i = \binom{\tau}{i} P_i'$. Analogously, we have $P_{b,i}'=\mathsf{\hat{P}_{b,SPC}}(im) -\sum_{j=1}^{i-1}\binom{i}{j} P_{b,j}'$ and $P_{b,i}=\binom{\tau}{i}P_{b,i}'$.
\end{exmp}

If $\nrs$ does not divide $\tau$, we must additionally consider the correlations between consecutive outer codewords to compute the exact error probabilities. Therefore, we adopt another approach where we set $n_\tau = \lceil\frac{\nrs}{\tau}\rceil$, for which \eqref{eq:Pf_general} becomes an upper bound and \eqref{eq:Pb_general} holds only approximately. However, simulation results show that for $\nrs\gg\tau$, both the \ac{FER} and \ac{BER} are accurately predicted by this approach.

\begin{exmp}[KP4 \& inner (16,15) SPC codes] \label{exmp:n16}
We have $\tau=3$ since three \ac{RS} symbols are covered by two \ac{SPC} codes. Since $\nrs$ is not an integer multiple of $\tau=3$, we have $n_\tau = \lceil\frac{\nrs}{\tau}\rceil = 182$. Numerical results show that \eqref{eq:Pf_general} and \eqref{eq:Pb_general} accurately predict the end-to-end \ac{FER} and \ac{BER} for low error rates.
\end{exmp}

\subsection{Error Rates with Uniform Symbol Interleaver} 
\label{sec:analysis:symbolinterleaver}
\begin{figure*}[!t]
    \centering
    \subfloat{\includegraphics{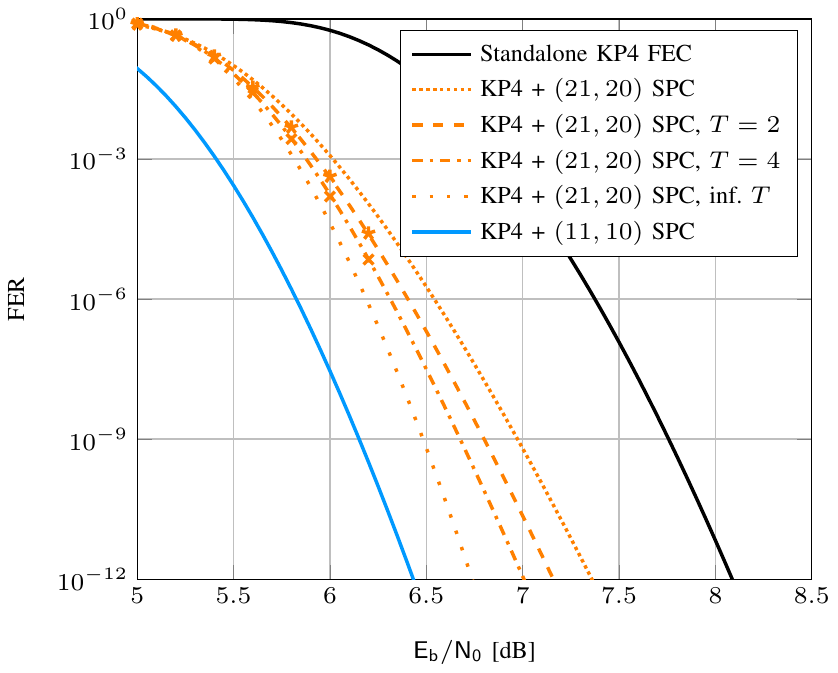}%
    \label{fig_first_case}}
    \hfil
    \subfloat{\includegraphics{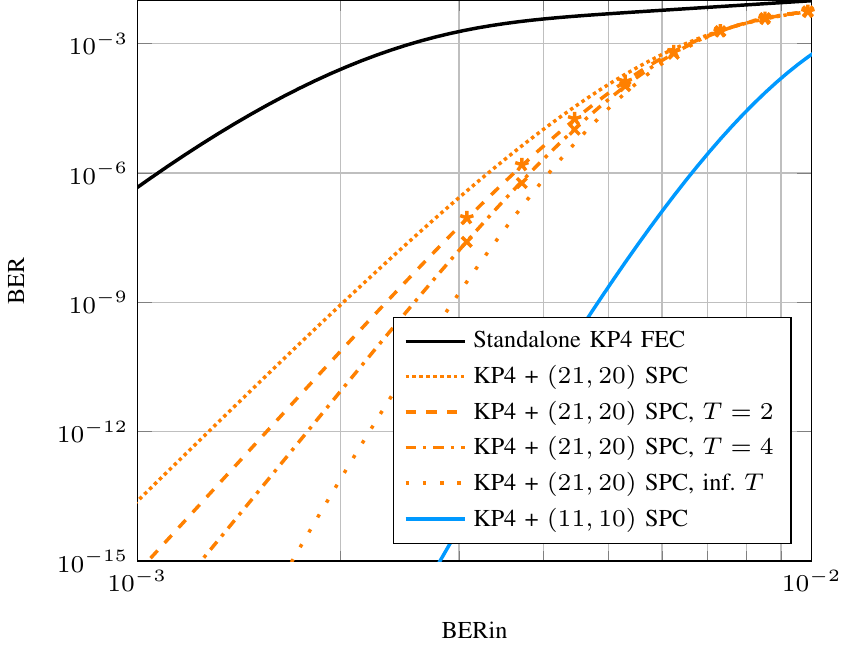}%
    \label{fig_second_case}}
    \caption{End-to-end \acp{FER} (left) and \acp{BER} (right) for \ac{BPSK} transmission over the AWGN channel, with and without interleaving between an outer KP4 code and an inner \ac{SPC} code of lengths $\nspc=11,21$.}
    \label{fig:cfec:rand_RS_interleaver}
\end{figure*}

Suppose now there is a uniform interleaver $\pia$ that randomly permutes the $\npia=T\cdot\nrs$ \ac{RS} symbols of $T$ outer \ac{RS} codewords.
If $T=1$, the error probabilities remain the same as without an interleaver since the error correction capability of the outer \ac{RS} code does not depend on the positions of the erroneous symbols.  

At the other extreme, in the limit $\npia\to\infty$, the symbol errors become uncorrelated after inner decoding and deinterleaving. Therefore, we must consider their bursty nature before interleaving to compute the symbol error probability after the deinterleaver. More precisely, the average symbol error rate and the average \acp{BER} after inner decoding and before deinterleaving are $\bar{P} = \frac{1}{\tau}\sum_{i=1}^\tau i P_i$ and $\bar{P}_b = \sum_{i=1}^\tau P_{b,i}=\Pbspc$, respectively.
The end-to-end \acp{FER} and \acp{BER} of the interleaved \ac{CatFEC} scheme are then obtained by inserting $\pt=\bar{P}$ and $p=\bar{P}_b$ into \eqref{eq:Pfrs}--\eqref{eq:Pbrs}.

For small $T$, as required by low-latency applications, the above asymptotic analysis is not precise enough. Instead, by carefully considering the correlations among the symbol errors at each decoding stage, we can express the \acp{FER} and \acp{BER} in closed form for the interleaved set-up. For $i=0,\ldots,\tau$, let $z_i$ be the total number of $\tau$-tuples with $i$ erroneous symbols within the $T$ outer codewords after inner Wagner decoding, and before deinterleaving and decoding.
The probability that the $T$ outer codewords contain in total $e=\sum_{i=1}^{\tau} i\cdot z_i$ individual symbol errors is then
\begin{align}
    P(e) = \sum\limits_{\substack{z_0,\ldots,z_{\tau}\,\geq 0 \\ \sum_{i=0}^{\tau} z_i \,=\, T\cdot n_\tau \\  \sum_{i=1}^{\tau} i\cdot z_i \,=\, e}} \binom{T\cdot n_\tau}{z_0,\,\ldots,\,z_{\tau}} \prod_{i=0}^{\tau} P_i^{z_i} \,.
    \label{eq:Pf_pi_finite_fixed_e}
\end{align}

The deinterleaver $\ipia$ distributes the $e$ symbol errors uniformly over the $T$ outer codewords but does not change $e$ itself. For $i=0,\ldots,T$, the joint probability that codeword $i$ will contain $0\leq e_i \leq \nrs$ symbol errors after deinterleaving is given by the multivariate hypergeometric distribution
\begin{align}
    g(e_0,\ldots,e_{T}) &= \frac{\prod_{i=1}^{T}\binom{\nrs}{e_i}}{\binom{T\nrs}{e}}
    \label{eq:hypergeom_pmf}
\end{align}
with $e=\sum_{i=1}^{T}e_i$. This can be seen by considering the (de)interleaving as a sampling without replacement. Consider an urn with $T\cdot\nrs$ balls of $T$ different colors, with $\nrs$ balls of each color. Suppose the balls from each color are labeled with numbers from one to $\nrs$. Each color represents one outer codeword and the balls their position within this codeword. By drawing $e$ balls, we select the symbol error positions after deinterleaving. The probability to pick $e_i$ balls from color $i$, $i=1,\ldots,T$, is then given by \eqref{eq:hypergeom_pmf}.

\ac{RS} \ac{BDD} of the $i$-th codeword will succeed if $e_i\leq t$, and fail if $e_i>t$. The overall end-to-end \ac{FER} of the interleaved \ac{CatFEC} scheme is
\begin{align}
    \Pf &= \sum\limits_{\substack{e_0,\ldots,e_{T}\,\geq 0}} \frac{\sum\limits_{i=1}^{T} \epsilon(e_i)}{T} \cdot g(e_0,\ldots,e_{T}) \cdot P\left(\sum_{i=0}^{T} e_i\right)
    \label{eq:Pf_pi_finite}
\end{align}
where $\epsilon(e_i) = 0$ if $e_i\leq t$ and $\epsilon(e_i) = 1$ if $e_i>t$.
Similarly, the end-to-end \ac{BER} is
\begin{align}
    \Pb &= \sum\limits_{\substack{e_0,\ldots,e_{T}\,\geq 0}}  \frac{\sum\limits_{i=1}^{T} e_i\cdot\epsilon(e_i)}{T\cdot\nrs} \cdot g(e_0,\ldots,e_{T}) \cdot P_b\left(\sum_{i=0}^{T} e_i\right)
    \label{eq:Pb_pi_finite}
\end{align}
where $P_b(e)$ is
\begin{align}
    \sum\limits_{\substack{z_0,\ldots,z_{\tau}\,\geq 0 \\ \sum_{i=0}^{\tau} z_i \,=\, T\cdot n_\tau \\  \sum_{i=1}^{\tau} i\cdot z_i \,=\, e}} \left(\frac{\frac{\sum\limits_{i=1}^{\tau} z_i \frac{P_{b,i}}{P_i} }{T\cdot n_\tau}}{\frac{e}{T\cdot\nrs}}\right)\cdot \binom{T\cdot n_\tau}{z_0,\,\ldots,\,z_{\tau}} \prod_{i=0}^{\tau} P_i^{z_i} \,.
    \label{eq:Pb_pi_finite_fixed_e}
\end{align}
Note that if $T\cdot\nrs$ does not divide $\tau$, then we can use the same trick as at the end of Sec.~\ref{sec:analysis:trivialinterleaver} and set $T\cdot n_\tau = \lceil\frac{T\cdot\nrs}{\tau}\rceil$ to compute \eqref{eq:Pf_pi_finite_fixed_e} and \eqref{eq:Pb_pi_finite_fixed_e}. As before, \eqref{eq:Pf_pi_finite_fixed_e} and \eqref{eq:Pf_pi_finite} are now upper bounds and \eqref{eq:Pb_pi_finite} and \eqref{eq:Pb_pi_finite_fixed_e} are approximations. As long as $T\cdot\nrs\gg\tau$, both $\Pf$ and $\Pb$ are closely approximated by \eqref{eq:Pf_pi_finite} and \eqref{eq:Pb_pi_finite}, respectively.

Fig.~\ref{fig:cfec:rand_RS_interleaver} plots the \ac{FER} $\Pf$ and \ac{BER} $\Pb$ with and without interleaving for the codes from Examples~\ref{exmp:n11} and \ref{exmp:n16}. The \ac{FER} and \ac{BER} of a standalone KP4 code are plotted in black as a benchmark. For $\nspc=11$, we have $\tau=1$, i.e., all symbol errors are uncorrelated before outer decoding, even without an interleaver. The \ac{FER} and \ac{BER}, therefore, remain the same in both cases and are shown by the solid blue line. 
For $\nspc=21$, we show the error probabilities without an interleaver (orange densely dotted), with uniform interleavers of lengths $T=2$ (orange dashed) and $T=4$ (orange dash-dotted), and an infinitely long uniform interleaver (orange loosely dotted). We see that $T=4$ suffices to achieve half of the asympototic $\EbNo$ interleaving gain at $\text{FER} = 10^{-12}$.
The orange markers depict simulation results for $\nspc=21$, $T=2,4$, and a fixed interleaver realization. The figure shows that the analysis describes the average ensemble performance and accurately predicts the \ac{FER} and \ac{BER} for specified (typical) interleaver realizations.

\section{Quantized Channel Outputs} 
\label{sec:analysis:coarslyquantized}
We extend the above analysis to coarsely quantized channel outputs. Consider a uniform $b$-bit quantizer with transfer function
\begin{align}
    f_\Delta(x) = \sgn(x) \cdot \Delta \cdot \min\bigg\lbrace 2^{b-1}-\frac{1}{2},\, \bigg\lfloor\frac{|x|}{\Delta}\bigg\rfloor+\frac{1}{2}\bigg\rbrace
\end{align}
where the quantization step $\Delta$ should be optimized. An input $x\in\mathbb{R}$ is quantized to $\lambda_i=(i-\frac{1}{2})\Delta$, $i=-2^{b-1}+1,\ldots,2^{b-1}$, if $g_{i-1}\leq x \leq g_i$, where
\begin{align}
    g_i= \begin{cases}
    -\infty, & i = -2^{b-1} \\
    i\Delta, &  -2^{b-1} < i < 2^{b-1} \\
    +\infty, & i =2^{b-1} \,. 
    \end{cases}
\end{align}
We derive the \ac{FER} for $\kspc=m$ where each \ac{RS} symbol is protected by a single \ac{SPC} codeword.

The channel \ac{LLR} $\bar{L}=f_\Delta(L)$ after quantization is distributed as
\begin{align}
    P_{\bar{L}}(\lambda_i) &= Q\left(\frac{g_{i-1}-\mu_\text{ch}}{\sigma_\text{ch}} \right) -  Q\left(\frac{g_i-\mu_\text{ch}}{\sigma_\text{ch}} \right)
\end{align}
for $i=-2^{b-1}+1,\ldots,2^{b-1}$. Apart from dealing with discrete \acp{LLR}, we must consider the event of multiple \acp{LRB} with equal reliability $\abs{\bar{L}}$. In this case, the Wagner decoder picks one of the \acp{LRB} uniformly at random and flips it if the parity check is not satisfied. Suppose there is only one erroneous bit with quantized \ac{LLR} $\bar{l}=\lambda_i$, $i\in \{ -2^{b-1}+1, \ldots, 0 \}$, which is also an \ac{LRB}.
Furthermore, let there be another $z$ \acp{LRB} with the same reliability but that are correct, i.e., with quantized \ac{LLR} $-\bar{l} = \lambda_{{-i}+1}$. This occurs with probability
\begin{align}
    \varphi_z(i) = P_{\bar{L}}(\lambda_i) P_{\bar{L}}(\lambda_{{-i}+1})^{z} \left( \sum\limits_{j=-i+2}^{2^{b-1}} P_{\bar{L}}(\lambda_j) \right)^{\nspc-z-1} \,.
\end{align}
There are $\binom{\nspc}{z+1}$ possibilities to arrange the $z+1$ \acp{LRB}, and the probability of picking and correcting the erroneous bit is $\frac{1}{z+1}$. Thus, the overall probability that the Wagner decoder corrects a single bit error is
\begin{align}
    \Pr{\mathcal{C}_1} &= \sum\limits_{i=-2^{b-1}+1}^{0}  \sum\limits_{z=0}^{\nspc-1} \binom{\nspc}{z+1}\frac{1}{z+1} \, \varphi_z(i)  \,.
    \label{eq:Pr_C1_quant}
\end{align}
Inserting \eqref{eq:Pr_Al} and \eqref{eq:Pr_C1_quant} into \eqref{eq:Pfspc} yields the \ac{SPC} \ac{FER} $\Pfspc$. The end-to-end \ac{FER} of the \ac{CatFEC} scheme follows from \eqref{eq:Pf_general} and $P_1=\Pfspc$.

\begin{figure}[t]
    \centering
	\includegraphics{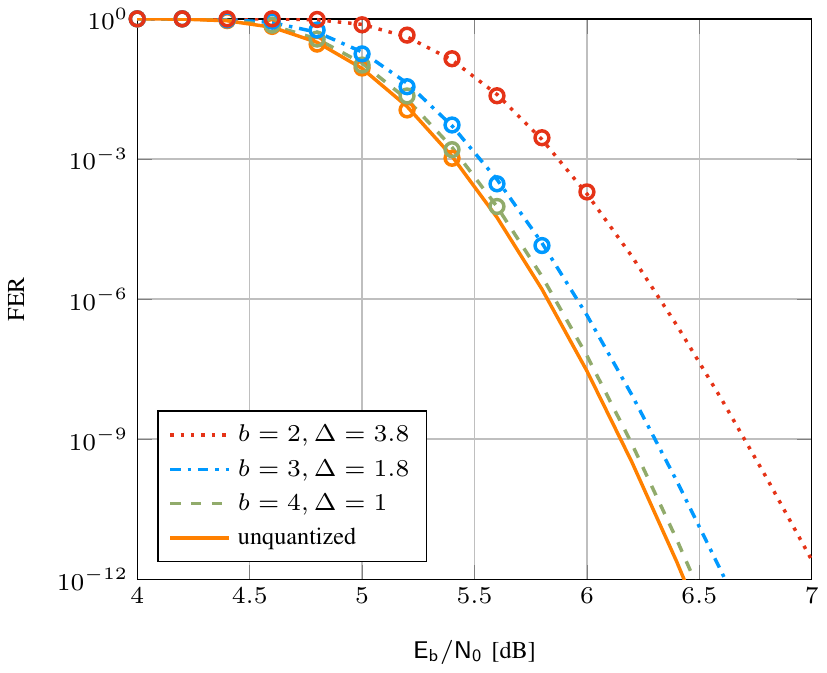}
    \caption{End-to-end \acp{FER} with inner (11,10) \ac{SPC} code for BPSK transmission over the AWGN channel when the channel output is quantized to $b=2,3,4$ bits.}
    \label{fig:error_analysis_quantized}
\end{figure}

Fig.~\ref{fig:error_analysis_quantized} plots the \ac{FER} of our \ac{CatFEC} scheme with an outer KP4 code and an inner \ac{SPC} code of length $\nspc=11$ when the channel output is quantized with $b=2$ (red dotted), $b=3$ (blue dash-dotted), and $b=4$ (green dashed) bits, and for an unquantized channel output (orange solid). The corresponding values of $\Delta$ are chosen heuristically and kept fixed for all $\EbNo$.
Again, lines show analytical error probabilities, and markers depict end-to-end \acp{FER} of simulations. Observe that $b=3$ bits resolution gives close-to-optimal performance even for low error rates.

\section{Extension to 4-ASK} 
\label{sec:analysis:higherorder}

The analysis extends to higher-order modulation and \ac{BMD} by considering the different reliabilities at each bit level. We derive the \ac{FER} of 4-ASK, i.e. $X \in\{\pm 3 \delta,\pm \delta\}$ where $\delta$ satisfies $\mathbb{E}[X^2]=1$. We consider $\kspc=m$ so that $\tau=1$ for simplicity.

We have $K=2$ and define $B_k=\chi_k(X)$, $k=1,2$, for a generic 4-ASK input $X$. The \acp{LLR} of the corresponding channel output $Y=y$ are
\begin{align}
    l_k = \log\frac{P_{B_k|Y}(0|y)}{P_{B_k|Y}(1|y)}, \quad k = 1,2 .
    \label{eq:Lch}
\end{align}
If the \acp{LLR} $L_k$ fulfill the symmetry constraint
\begin{equation}
    p_{L_k|B_k}(l|0) = p_{L_k|B_k}(-l|1), \quad l\in\mathbb R,\; k=1,2
    \label{eq:def_symmetry_channel}
\end{equation}
then we may assume that an all-zeros codeword was transmitted. However, the bit channels $p_{L_k|B_k}(.)$ are generally not symmetric.

To introduce symmetrized counterparts for 4-ASK, we use \emph{channel adapters} \cite{hou_capacity-approaching_2003} that apply a pseudo-random binary scrambling string at both the transmitter and receiver. As a result, $L_k$ is replaced with
\begin{equation}
\tilde L_k = L_k \cdot (1-2B_k)\label{eq:scrambling}
\end{equation}
and the bit-channels $p_{\tilde L_k|B_k}(.)$ are symmetric since
\begin{equation}
    p_{\tilde L_k|B_k}(l|0) = p_{\tilde L_k|B_k}(-l|1).
\end{equation}
We also use \emph{surrogate channels} \cite{peng2006,franceschini2006,sason_universal_2009,steiner2016protograph} and approximate the $p_{\tilde L_k|B_k}(.)$ by \ac{AWGN} channels with uniform binary inputs to simplify calculating \acp{CDF}. We require the actual channel and its surrogate to have the same uncertainty. Let the surrogate be $\breve Y_k = \breve X_k + \breve N_k$ with $\breve X_k \in \{-1,+1\}$ and $\breve N_k \sim \mathcal{N}(0,\breve \sigma_k^2)$ for $k=1,2$. For each \ac{SNR}, we compute the channel parameters
\begin{align}
\breve\sigma_k^2: H(\breve B_k|\breve Y) = H(B_k|Y), \quad k=1,2
\end{align}
where $H(X|Y)$ is the average entropy of $X$ given $Y$. Under the Gaussian approximation, the \acp{LLR} $\tilde{L}_k$ are Gaussian with mean $\breve\mu_{\text{ch},k} = 2\big/{\breve\sigma_k^2}$ and variance $\breve\sigma_{\text{ch},k}^2 = 4\big/{\breve\sigma_k^2}$, and the hard-decision \ac{BER} is $p_k = Q\left( { \breve\mu_{\text{ch},k}}\big/{ \breve\sigma_{\text{ch},k}} \right)$ for $k=1,2$.

Suppose $N_1$ code bits of an \ac{SPC} codeword are mapped to bit level $B_1$, and the other $N_2=\nspc-N_1$ code bits are mapped to bit level $B_2$. The \ac{SPC} \ac{FER} after Wagner decoding is
\begin{align}
    \Pfspc(N_1,N_2) &=   \sum\limits_{\substack{0 \leq \ell_1 \leq N_1 \\  0 \leq \ell_2 \leq N_2 \\ \ell_1+\ell_2\geq 1}} \Pr{\mathcal{A}_{\ell_1,\ell_2}} - \Pr{\mathcal{C}_{1}}
\end{align} 
where 
\begin{align}
    \Pr{\mathcal{A}_{\ell_1,\ell_2}} &= \prod_{k=1}^2 \binom{N_k}{\ell_k} p_k^{\ell_k} (1-p_k)^{N_k-\ell_k} \\
    \Pr{\mathcal{C}_{1}} &=  \sum_{k=1}^2 N_k \int\limits_{-\infty}^{0} \phi_k(a) \,\diff a 
\end{align}
with
\begin{align}
    \phi_1(a) &= p_{\tilde{L}_1}(a)\; Q\left(\tfrac{-a-\breve\mu_{\text{ch},1}}{\breve\sigma_{\text{ch},1}} \right)^{N_1-1} Q\left(\tfrac{-a-\breve\mu_{\text{ch},2}}{\breve\sigma_{\text{ch},2}} \right)^{N_2}  \\
    \phi_2(a) &= p_{\tilde{L}_2}(a)\; Q\left(\tfrac{-a-\breve\mu_{\text{ch},1}}{\breve\sigma_{\text{ch},1}} \right)^{N_1} Q\left(\tfrac{-a-\breve\mu_{\text{ch},2}}{\breve\sigma_{\text{ch},2}} \right)^{N_2-1} \,.
\end{align}

We assume the \ac{SPC} code bits are alternately mapped to the two bit levels $B_1$ and $B_2$. This leads to correlated \ac{SPC} frame errors and, since $\tau=1$, also to correlated \ac{RS} symbol errors of length $\tau' = \frac{\mathrm{LCM}(\nspc,2)}{2}$. Let $P_i$ be the probability that exactly $i$ RS symbols of the $\tau'$-tuple are in error before outer decoding for $i=0,\ldots,\tau'$. Let $z_i$ be the total number of such $\tau'$-tuples with $i$ erroneous symbols within one RS codeword. We have $\sum_{i=0}^{\tau'} P_i = 1$ and $\sum_{i=0}^{\tau'} z_i= n_{\tau'}$. 

For example, for an outer KP4 code we have $\nspc=m+1=11$, $\tau'=2$ and $n_{\tau'}=\nrs/2$. The \acp{FER} of the two \ac{SPC} codewords within one $2$-tuple are
\begin{align}
    \Pfspc_1 & = \Pfspc(\lceil\tfrac{\nspc}{2}\rceil,\lfloor\tfrac{\nspc}{2}\rfloor) \\
    \Pfspc_2 & = \Pfspc(\lfloor\tfrac{\nspc}{2}\rfloor,\lceil\tfrac{\nspc}{2}\rceil)
\end{align}
yielding
\begin{align}
    P_1 &= (1-\Pfspc_1) \Pfspc_2 + \Pfspc_1 (1-\Pfspc_2) \label{eq:P1_PAM4}\\
    P_2 &=  \Pfspc_1 \Pfspc_2 \,.
\end{align}
The overall end-to-end \ac{FER} of the \ac{CatFEC} scheme is thus
\begin{align}
    \Pf = \sum\limits_{\substack{z_0,\ldots,z_{\tau'}\,\geq 0 \\ \sum_{i=0}^{\tau'} z_i \,=\, n_{\tau'} \\  \sum_{i=1}^{\tau'} i\cdot z_i \,\geq\, t+1}} \binom{n_{\tau'}}{z_0,\,\ldots,\,z_{\tau'}} \prod_{i=0}^{\tau'} P_i^{z_i} \,.
    \label{eq:Pf_higer_order_mod}
\end{align}

\begin{figure}[t]
    \centering
   	\includegraphics{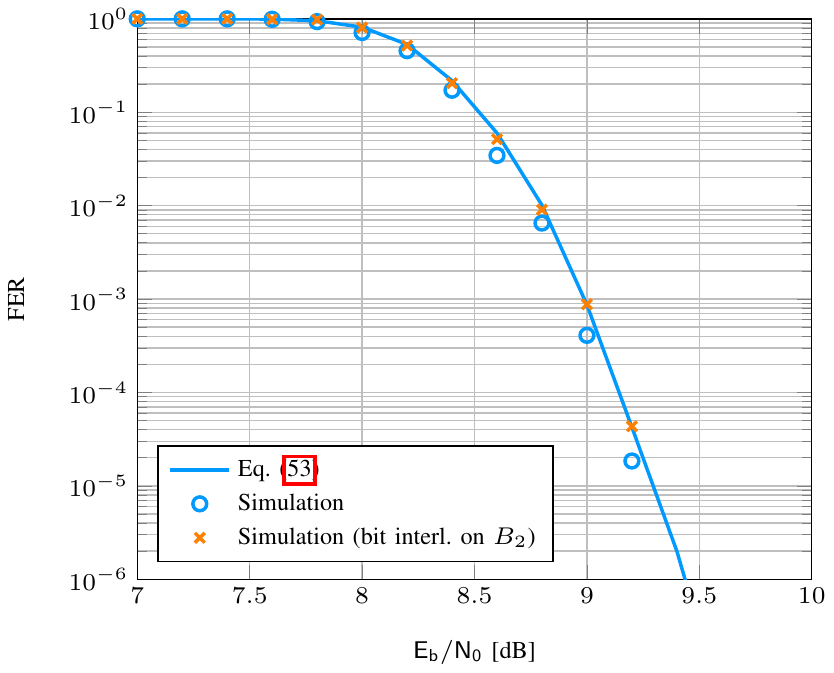}
    \caption{End-to-end \acp{FER} with inner (11,10) \ac{SPC} code for 4-ASK transmission over the AWGN channel.}
    \label{fig:error_analysis_4ask}
\end{figure}
Fig.~\ref{fig:error_analysis_4ask} shows the approximate end-to-end \ac{FER} from \eqref{eq:Pf_higer_order_mod} for the \ac{CatFEC} scheme with an inner $(11,10)$ \ac{SPC} code and 4-ASK transmission over the \ac{AWGN} channel (solid blue line). We also plot numerical simulation results (blue circles) obtained with the alternating bit-to-symbol mapping described above. As can be seen, the approximations of the surrogate channel framework introduce a small discrepancy between the predicted and observed \acp{FER}. However, the $\EbNo$ gap is almost negligible. We further provide numerical results (orange marks) of the \ac{CatFEC} scheme with an alternating bit-to-symbol mapping, but where a random bit interleaver of length $\abs{\pi_{\mathsf{bit}}} = \nrsb\cdot \frac{\nspc}{\kspc}\cdot\frac{1}{2} =2992$ permutes the code bits mapped to bit level $B_2$. Such an interleaver breaks the correlation of consecutive bits but preserves the bit reliabilities within the \ac{SPC} codewords. The simulated \acp{FER} are accurately predicted by \eqref{eq:Pf_higer_order_mod}, which suggests that the main source of inaccuracy is due to the correlations between the bit levels that are ignored by the parallel surrogate channel.

Alternatively, one could use \ac{MLC} \cite{imai1977new}, \cite{wachsmann1999multilevel}, where a separate \ac{SPC} code protects each bit level. This leads again to $n_{\tau'}=\nrs/2$ independent 2-tuples of \ac{RS} symbols mapped to the two bit levels $B_1$ and $B_2$. The end-to-end \ac{FER} of the \ac{CatFEC} scheme with \ac{MLC} and \ac{BMD} is then obtained by inserting $\Pfspc_1 = \Pfspc(\nspc,0)$ and $\Pfspc_2 = \Pfspc(0,\nspc)$ into \eqref{eq:P1_PAM4}--\eqref{eq:Pf_higer_order_mod}.
In a more general setting, \ac{SPC} codes of different lengths can be used for each bit level, giving another degree of freedom to optimize the \ac{CatFEC} system.

\begin{figure*}[!t]
    \centering
    \subfloat{\includegraphics{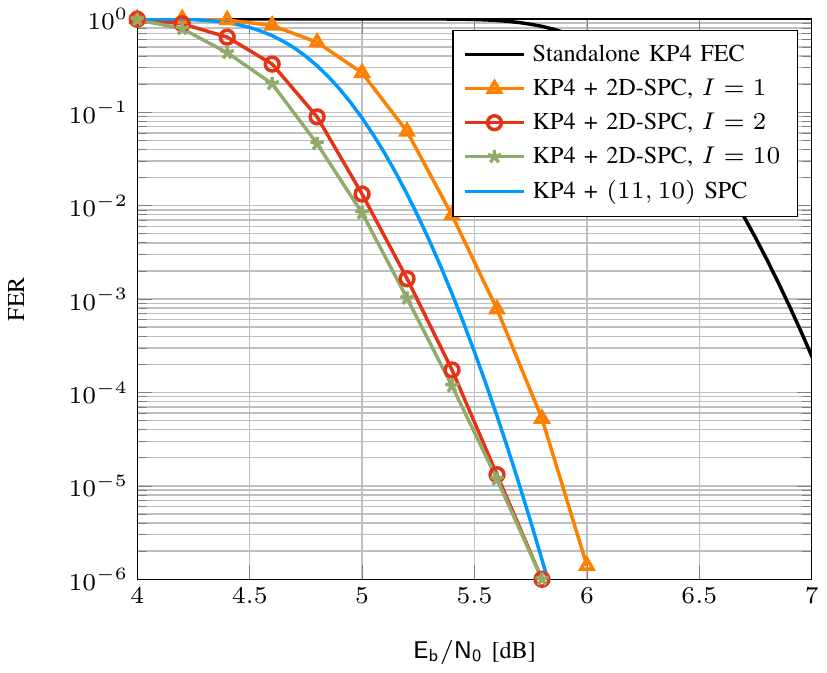}}%
    \hfil
	\subfloat{\includegraphics{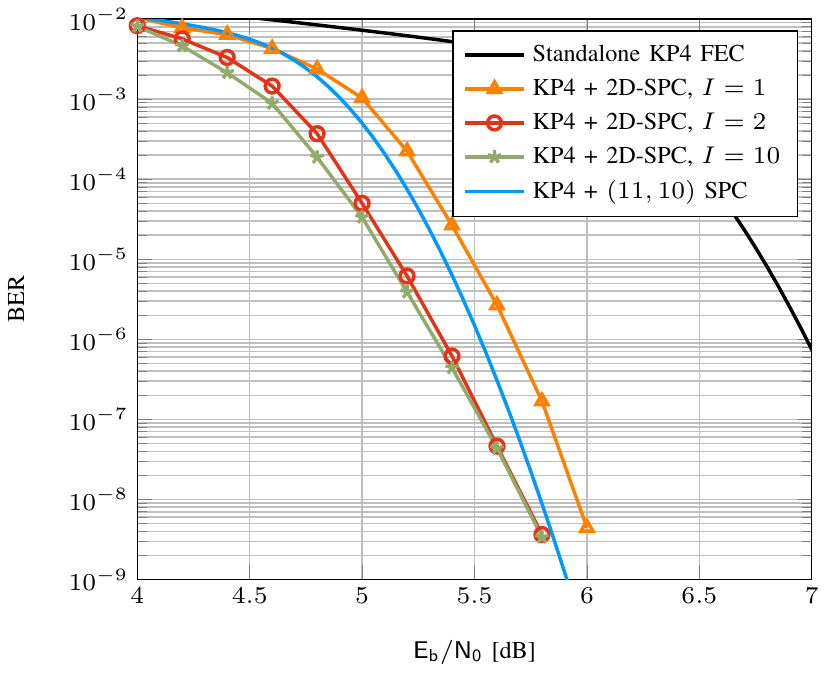}}%
    \caption{End-to-end \acp{FER} (left) and \acp{BER} (right) of a \ac{CatFEC} scheme with an outer KP4 code and a) an inner $(21^2,20^2)$ \ac{2D-SPC} code with soft product decoding, b) an inner $(11,10)$ SPC code and Wagner decoding. The plots were generated for \ac{BPSK} transmission over the AWGN channel without interleaving.}
    \label{fig:concat:fer:2d-spc}
\end{figure*}

\section{Comparison to Alternative Schemes}
\label{ch:comparison}
Consider \ac{BPSK} transmission over the \ac{AWGN} channel. We compare the performance of our \ac{CatFEC} scheme to two alternative \ac{CatFEC} schemes where the inner \ac{SPC} code is replaced by a \ac{2D-SPC} code or a $(128,120)$ extended Hamming code. We again apply a soft-in, hard-out inner decoder and use the KP4 outer code. 

We use Monte Carlo simulations to evaluate the performance. When we can analytically compute the \ac{FER} and \ac{BER}, we plot the error probabilities using solid lines without markers. Simulation results are plotted with markers. We adjust the length of the inner \ac{SPC} code to obtain approximately the same end-to-end rate for each \ac{CatFEC} scheme. We assume separate \ac{FEC} per optical signal and, therefore, no interleaving for all simulations.

\subsection{Inner (441,400) 2D-SPC Code}
\label{sec:2dspc}
Consider a two-dimensional product code with $(21,20)$ SPC component codes, i.e., the inner code parameters are $(21^2,20^2)$. The \ac{2D-SPC} code rate is $0.9070$, and the end-to-end transmission rate is $R=0.8570$ ($16.68\%$ \ac{OH}). We compare to the proposed \ac{CatFEC} scheme with an inner $(11,10)$ \ac{SPC} code where the end-to-end transmission rate is $R=0.8590$ ($16.42\%$ OH). We apply soft product decoding with bitwise \ac{MAP} decoding of the component SPC codes and with hard decisions after a maximum of $I$ full iterations. After each full iteration, the decoder outputs the hard decisions if all \ac{SPC} constraints are satisfied.

The \ac{FER} and \ac{BER} of the \ac{2D-SPC} code are depicted in Fig.~\ref{fig:concat:fer:2d-spc} for $I = 1,2,10$ decoding iterations. Note that $I = 2$ achieves almost the full coding gain for \acp{FER} below $10^{-5}$. This observation agrees with \cite{rankin2001}, where the performance of $n$-dimensional SPC product codes converges after $n$ decoding iterations. Next, the \ac{FER} slope of the \ac{CatFEC} scheme with the inner \ac{2D-SPC} code and soft product decoding is smaller than that of the proposed \ac{CatFEC} scheme. While at $\mathrm{FER} = 10^{-2}$ the former code gains approximately $\SI{0.25}{dB}$ over latter code, the \ac{FER} curves of both schemes intersect at $\mathrm{FER} \approx 10^{-6}$, making the inner \ac{SPC} code preferable for low error rates. Both effects are also observed for the \acp{BER}.

\subsection{Inner (128,120) Extended Hamming code} 
\label{sec:Hamming}
\begin{figure*}[t]
    \centering
    \subfloat{\includegraphics{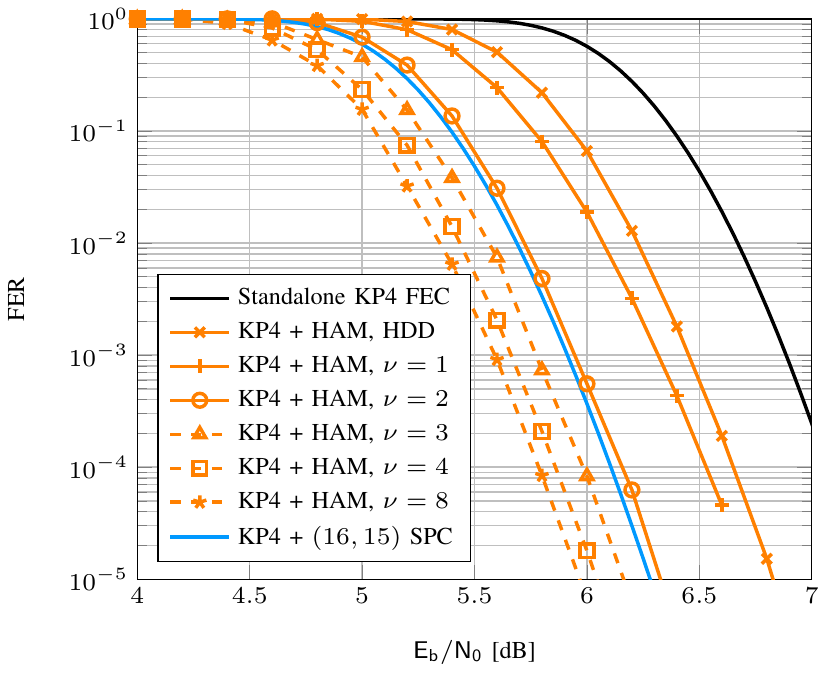}}%
    \hfil
    \subfloat{\includegraphics{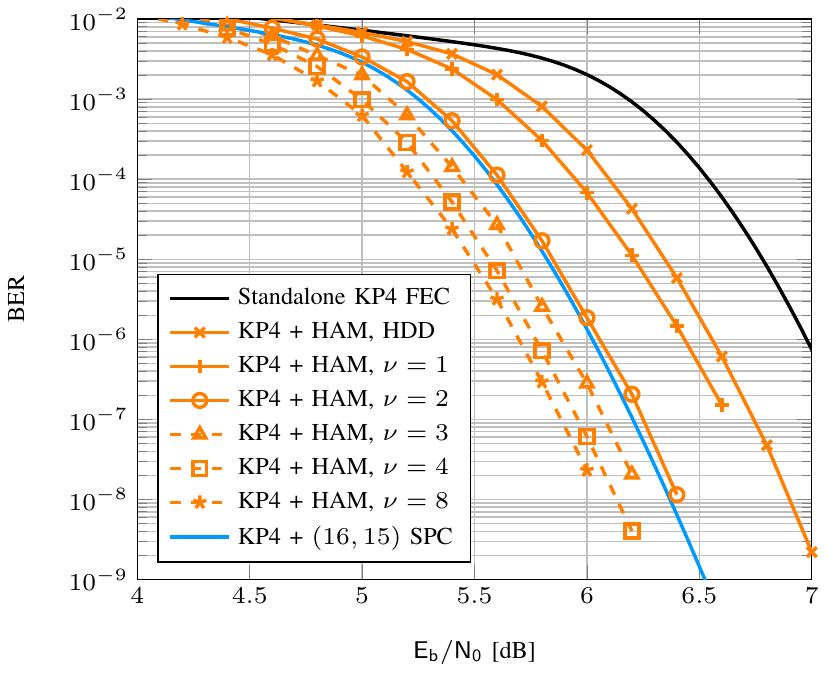}}
    \caption{End-to-end \acp{FER} (left) and \acp{BER} (right) of a \ac{CatFEC} scheme with an outer KP4 code and a) an inner $(128,120)$ Hamming code with Chase decoding, b) an inner $(16,15)$ SPC code and Wagner decoding. The plots were generated for \ac{BPSK} transmission over the AWGN channel without interleaving.}
    \label{fig:concat:fer:Hamming}
\end{figure*}
Consider now the $(128,120)$ extended Hamming code as an inner code. The end-to-end transmission rate of the concatenated system is $R=0.8858$ ($12.89\%$ OH). 
We compare this \ac{CatFEC} scheme to the proposed scheme with the $(16,15)$ SPC code of the same rate as the Hamming code. 

The Hamming code is decoded with a Chase decoder \cite[Alg.~2]{chase1972} with $2^\nu$ test patterns, where $\nu$ is the number of \ac{LRB} positions used to form the test list.
Given the input \acp{LLR} $\boldsymbol{l}$, the Chase decoder first computes the hard decisions $\boldsymbol{d}=\chi(\sgn(\boldsymbol{l}))$ and reliability values $\boldsymbol{r}=|\boldsymbol{l}|$, where both the sign and absolute value are taken element-wise. Based on $\boldsymbol{r}$, the decoder identifies the positions of the $\nu$ \acp{LRB} and forms a test list $\mathcal{T}_\nu$ by making $2^\nu$ copies of the vector $\boldsymbol{d}$ and replacing the $\nu$ \acp{LRB} by all binary combinations of length $\nu$. We call each member $\boldsymbol{z}\in\mathcal{T}_\nu$ of the list a test word, and the vector $\boldsymbol{t} = \boldsymbol{d}\oplus\boldsymbol{z}$ the corresponding test pattern. For each test word $\boldsymbol{z}$ in the test list, one \ac{BDD} attempt is performed. The decoder returns the binary error vector $\boldsymbol{e}$ (with respect to $\boldsymbol{z}$) if successful, and otherwise it flags a decoding failure. For each successful \ac{BDD} attempt, the Chase decoder computes the \emph{analog weight} as the inner product
\begin{align}
    w = \sum_{i=1}^n \tilde{e}_i r_i
    \label{eq:analog_weight}
\end{align}
of the combined error vector $\boldsymbol{\tilde{e}} = \boldsymbol{t}\oplus\boldsymbol{e}$.
The Chase decoder keeps track of the combined error vector $\boldsymbol{\tilde{e}}^*$ of lowest analog weight $w^*$ and outputs $\boldsymbol{\hat{c}} = \boldsymbol{d}\oplus\boldsymbol{\tilde{e}}^*$.

Chase decoding as in \cite[Alg.~2]{chase1972} guarantees correcting up to $d-1$ errors for a code with minimum distance $d$ by considering only the $\nu = \lfloor\frac{d}{2}\rfloor$ \acp{LRB}.  The performance improves by choosing larger values of $\nu$. 
Since the minimum distance of the extended Hamming code is $d=4$, we have $\nu = \lfloor\frac{d}{2}\rfloor = 2$, i.e., $2^{\lfloor\frac{d}{2}\rfloor} = 4$ test patterns. However, we also study larger test list sizes to illustrate the potential gains.

\subsection{Decoding Complexity} 
\label{sec:decoding-complexity}
We quantify the complexity by the number of elementary operations required to decode one codeword, i.e., the number of logical XOR and AND operations, as well as the number of additions of two reals. We omit to count the number of pairwise comparisons of reals, as this number depends on the sorting algorithm. For example, Wagner decoding of an SPC code of length $\nspc$ is particularly simple: $\nspc-1$ bitwise XOR operations for the syndrome computation and one XOR to potentially flip the \ac{LRB}.

The complexity of Chase decoding a $(\nh,\kh)$ Hamming code depends on the number $\nu$ of \acp{LRB} used to generate the test list. For each of the $2^{\nu}$ test words, we need to compute the syndrome and the analog weight of the combined error vector. We use a syndrome-based \ac{LUT} implementation of \ac{BDD} and neglect the complexity of the look-up operation. The binary syndrome vector is computed by multiplying with the parity-check matrix of dimension $(\nh-\kh)\times\nh$, which in general requires up to $(\nh-\kh)\nh$ AND and $(\nh-\kh)(\nh-1)$ XOR operations. Each analog weight calculation requires $\nh$ XOR operations to determine the combined error vector $\boldsymbol{\tilde{e}}$, as well as $\nh-1$ real additions for the inner product \eqref{eq:analog_weight}. This inner product does not require real multiplications since the entries of $\boldsymbol{\tilde{e}}$ are either zero or one. The final calculation of $\boldsymbol{\hat{c}}$ requires another $\nh$ XOR operations.

In contrast, \ac{HDD} of the Hamming code requires $(\nh-\kh)\nh$ AND and $(\nh-\kh)(\nh-1)$ XOR operations to compute the syndrome, and another $\nh$ XOR operations to add the error vector from the \ac{LUT} to the input. Note that setting $\nu=0$ in the above complexity calculations yields a larger complexity score since some steps like the computation of $\boldsymbol{\tilde{e}}$ and \eqref{eq:analog_weight} are not required for HDD.\footnote{\ac{HDD} algorithms require less power and input/output bandwidth than \ac{SDD} algorithms such as Wagner and Chase decoding.}

\begin{table}
    \caption{Number of elementary operations required for a) Wagner decoding of eight $(16,15)$ SPC codewords, b) Chase decoding of one $(128,120)$ Hamming codeword with $\nu=1,2,3,4,8$.}
	\label{tab:concat:decoding_complexity_example}
    \centering
	\begin{tabular}{l|c|c|c}
		&XORs & ANDs & real ADDs \\
		\hline
		$8\times$ SPC(16,15) &128& -- & -- \\
		HAM(128,120), HDD &1144& 1024 & -- \\
		HAM(128,120), $\nu=1$ & 2416 & 2048 & 254 \\
		HAM(128,120), $\nu=2$ & 4704 & 4096 & 508 \\
		HAM(128,120), $\nu=3$ & 9280 & 8192 & 1016 \\
		HAM(128,120), $\nu=4$ & 18432 & 16384 & 2032 \\
		HAM(128,120), $\nu=8$ & 292992 & 262144 & 32512 
	\end{tabular}
\end{table}

To make the comparison as fair as possible, we compute the complexity of decoding the same number of code bits or, equivalently, the same number of information bits. Table~\ref{tab:concat:decoding_complexity_example} shows the complexity of decoding eight codewords of the $(16,15)$ SPC code and one codeword of the $(128,120)$ Hamming code. Even for \ac{HDD}, the Hamming code requires one order of magnitude more logical operations than Wagner decoding of the \ac{SPC} codes. For $\nu=2$, the complexity is almost two orders of magnitude larger. For larger $\nu$, the Hamming decoder can become prohibitively complex.

Fig.~\ref{fig:concat:fer:Hamming} shows the end-to-end \acp{FER} (left) and \acp{BER} (right) of the \ac{CatFEC} schemes for \ac{BPSK} and transmission over the \ac{AWGN} channel. 
Remarkably, the inner $(16,15)$ SPC code slightly outperforms the Hamming code in both \ac{FER} and \ac{BER} even when $\nu=\lfloor\frac{d}{2}\rfloor=2$. The \ac{SPC} code exhibits a coding gain of $\sim\!\SI{0.5}{dB}$ compared to \ac{HDD} the Hamming code. Chase decoding with $\nu>2$ lets the Hamming code improve on the \ac{SPC} code but with substantial decoding complexity.

\section{Unit-Memory ISI Channels}
\label{sec:ISI_channel}

\begin{figure*}[!t]
    \centering
    \includegraphics{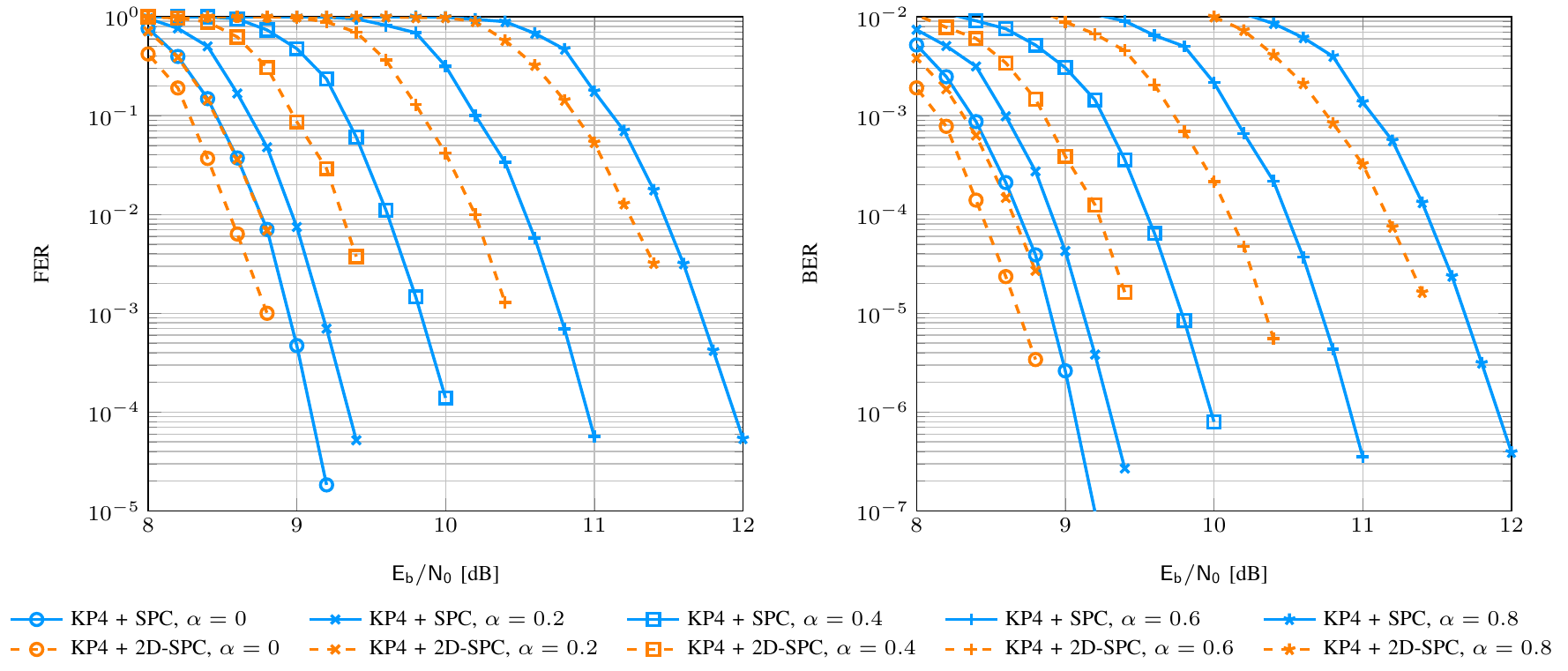}
    \caption{End-to-end \acp{FER} (left) and \acp{BER} (right) of \ac{CatFEC} schemes with an inner $(11,10)$ SPC code and an inner $(441,400)$ \ac{2D-SPC} code ($I=2$). The \acp{FER} and \acp{BER} are shown for the unit-memory \ac{ISI} channel with 4-ASK modulation and interference levels $\alpha=0, 0.2, 0.4, 0.6, 0.8$, and without interleavers.}
    \label{fig:concat:PAM4_ISI:2D-SPC:trivial_PIa}
\end{figure*}
\begin{figure*}[t]
    \centering
    \includegraphics{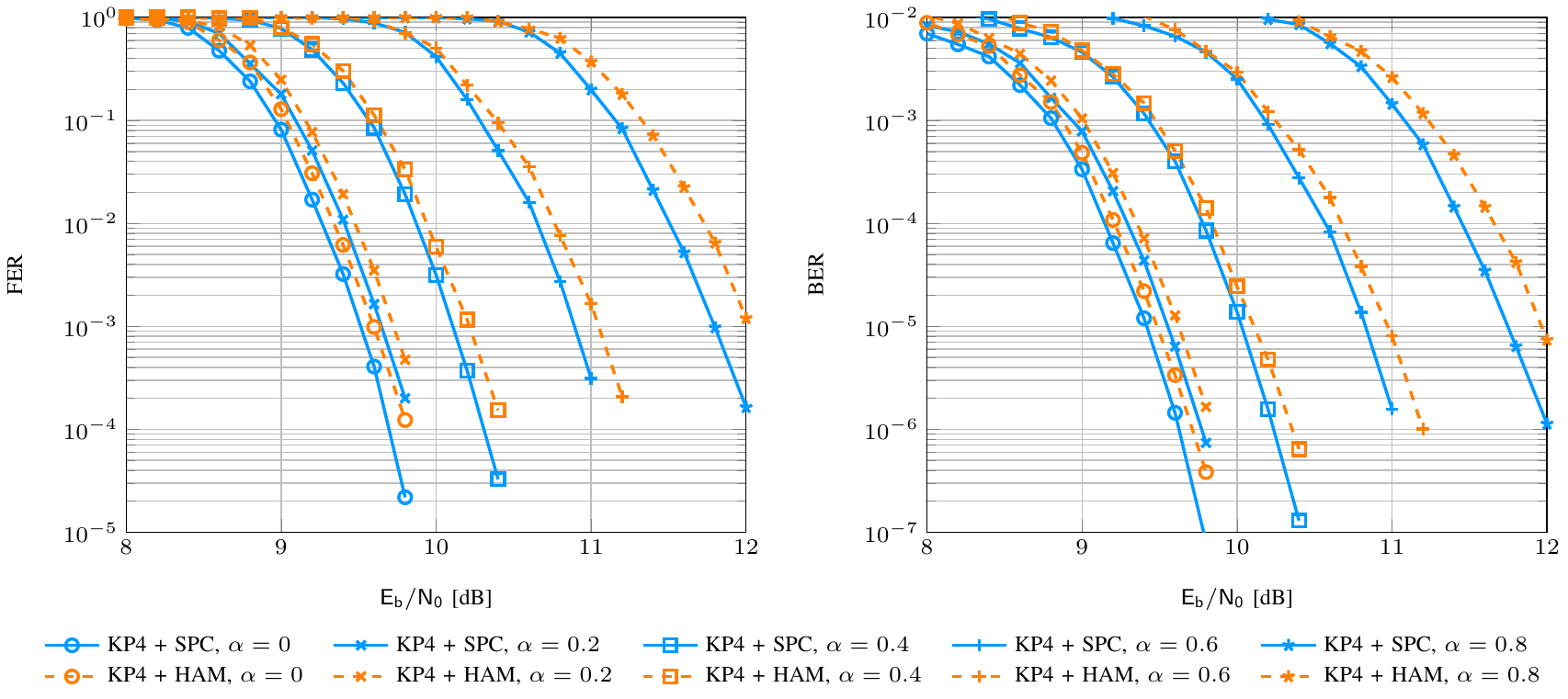}
    \caption{End-to-end \acp{FER} (left) and \acp{BER} (right) of \ac{CatFEC} schemes with an inner $(16,15)$ SPC code and an inner $(128,120)$ Hamming code ($\nu=2$). The \acp{FER} and \acp{BER} are shown for the unit-memory \ac{ISI} channel with 4-ASK modulation and interference levels $\alpha=0, 0.2, 0.4, 0.6, 0.8$, and without interleavers.}
    \label{fig:concat:PAM4_ISI:HAM:trivial_PIa}
\end{figure*}

A common model in optical communications is the unit-memory \ac{ISI} channel
\begin{align}
     Y_j = X_j + \alpha X_{j-1} + N_j
     \label{eq:ISI-channel}
\end{align}
with interference level $0\leq\alpha\leq 1$ and $N_j \sim\mathcal{N}(0,\sigma^2)$.
The model~\eqref{eq:ISI-channel} corresponds to an effective channel after applying \ac{LMMSE} filtering and noise whitening. 
In this section, we consider 4-ASK modulation with Gray labeling \cite{gray1953} and $\mathbb{E}[X_j^2]=1$ for all $j$.

We plot the error rates against $\EbNo = \frac{\SNR}{2 R}$, where $R$ is the end-to-end transmission rate and $\SNR = \frac{1+\alpha^2}{\sigma^2}$. 
We use the consecutive bit-to-symbol mapping described in Sec.~\ref{sec:analysis:higherorder} with \ac{BMD} at the receiver. Given a channel output vector $\boldsymbol{y}$, the demapper computes the bit-wise posterior \acp{LLR} in two steps.
\begin{enumerate}
    \item Compute the symbol-wise posterior probabilities $P_{X_j|\boldsymbol{Y}}(x|\boldsymbol{y})$ via the forward-backward algorithm \cite{bcjr}.
    \item Compute the posterior \ac{LLR} of the $k$-th bit $\chi_k(X_j)$ of the $j$-th channel input symbol $X_j$ via marginalization:
        \begin{align}
            l_{j,k} &= \log_2\frac{\sum\limits_{\tilde{x}\in\mathcal{X}\,:\,\chi_k(\tilde{x})=0} P_{X_j|\boldsymbol{Y}}(\tilde{x}|\boldsymbol{y})}{\sum\limits_{\tilde{x}\in\mathcal{X}\,:\,\chi_k(\tilde{x})=1} P_{X_j|\boldsymbol{Y}}(\tilde{x}|\boldsymbol{y})} \,.
        \end{align}
\end{enumerate}

Consider the two codes from Section~\ref{sec:2dspc}.
Fig.~\ref{fig:concat:PAM4_ISI:2D-SPC:trivial_PIa} shows the \acp{FER} and \acp{BER} for interference levels $\alpha=0, 0.2, 0.4, 0.6, 0.8$ (from left to right). The error rates for the \ac{CatFEC} scheme with the inner $(21^2,20^2)$ \ac{2D-SPC} code and $I = 2$ soft product iterations are depicted by dashed orange lines. The solid blue lines show the error rates for the \ac{CatFEC} scheme with the inner $(11,10)$ SPC code and Wagner decoding. 

Consider $\alpha=0$ (circles), i.e., 4-ASK transmission over the interference-free channel. The \ac{2D-SPC} code again has a slight advantage over the (11,10) \ac{SPC} code at high \acp{FER}. However, the \ac{SPC} code curve is steeper, which suggests that this code performs better at lower error rates. 
As $\alpha$ increases to 0.6, the gap between the \ac{2D-SPC} and \ac{SPC} curves increases slightly. Interestingly, the gap reduces again as $\alpha$ increases to 0.8. All effects can be observed for the \acp{FER} and \acp{BER}.

The performance of the two codes from Sec.~\ref{sec:Hamming} is similar. Fig.~\ref{fig:concat:PAM4_ISI:HAM:trivial_PIa} shows the \ac{FER} and \ac{BER} for interference levels $\alpha=0, 0.2, 0.4, 0.6, 0.8$ (from left to right). The error rates of the \ac{CatFEC} scheme with the inner $(128,120)$ Hamming code and Chase decoding with $\nu=2$ are depicted with dashed orange lines. The solid blue lines show the error rates with the inner $(16,15)$ SPC code and Wagner decoding. The \ac{SPC} code gain slightly increases with the interference level compared to the Hamming code.

For $\alpha=0$, the results are consistent with their \ac{BPSK} counterparts, which shows that the \ac{BPSK}+\ac{AWGN} model is a reasonable proxy for code design. The analysis in Sec.~\ref{ch:analysis} is thus helpful to evaluate error rates without computation-heavy simulations, even when using realistic interleavers. 

We conclude with two remarks. First, in the presence of \ac{ISI}, the performance of all presented \ac{CatFEC} schemes can be further improved by using a channel interleaver, see Sec.~\ref{ch:CFEC}. We expect the shorter \ac{SPC} codes to require shorter channel interleavers than, e.g., the (128,120) Hamming code. 
Second, there is potential to reduce the computational complexity of Step 1) of the \ac{LLR} computation (forward-backward algorithm) when using \ac{SPC} codes since we do not require the soft information to be true \acp{LLR}. 

\section{Conclusion}
\label{ch:conclusion}

We investigated \ac{FEC} codes for short-reach fiber optic links with strict latency requirements. We proposed a \ac{CatFEC} scheme with an outer KP4 code and inner \ac{SPC} codes, and we provided analytical expressions for the end-to-end \ac{FER} and \ac{BER} without and with \ac{RS} symbol interleavers. Simulations show that \ac{SPC} codes as inner codes achieve similar error rates with considerably lower decoding complexity as \ac{CatFEC} schemes with more sophisticated inner codes. The relative coding gains for \ac{BPSK} transmission over \ac{AWGN} channels and for 4-ASK signaling over \ac{ISI} channels with \ac{AWGN} are similar, which justifies using the former as a proxy for code design for the latter.

\bibliographystyle{IEEEtran}
\bibliography{IEEEabrv,confs-jrnls,literature-astro}

\end{document}